\documentclass[12pt]{article}
\usepackage{enumerate}
\usepackage{amsfonts}
\usepackage{amsmath}
\usepackage{amssymb}
\usepackage{amsthm}

\def\H{\mathcal{H}}

\def\S{\mathfrak{S}}
\def\F{\mathfrak{F}}
\def\C{\mathfrak{C}}
\def\T{\mathfrak{T}}
\def\B{\mathfrak{B}}

\newcommand{\supp}{\mathrm{supp}}
\newcommand{\rank}{\mathrm{rank}}
\newcommand{\id}{\mathrm{Id}}
\newcommand{\Tr}{\mathrm{Tr}}

\newcommand{\shs}{\hspace{1pt}}

\newcounter{defin}  \newcounter{lemma}  \newcounter{theorem}
\newcounter{property} \newcounter{corol}  \newcounter{remark} \newcounter{example}

\newenvironment{lemma}{\par\refstepcounter{lemma}
     \textbf{Lemma \thelemma.} }{\rm\par}
\newenvironment{theorem}{\par\refstepcounter{theorem}
     \textbf{Theorem \thetheorem.}\ }{\rm\par}

\newenvironment{remark}{\par\refstepcounter{remark}
     \textbf{Remark \theremark.}}{\rm\par}
\newenvironment{example}{\par\refstepcounter{example}
     \textbf{Example \theexample.}}{\rm\par}

\begin{document}

\title{Uniform finite-dimensional approximation of basic capacities of energy-constrained channels}
\author{M.E. Shirokov\footnote{Steklov Mathematical Institute, RAS, Moscow, email:msh@mi.ras.ru}}
\date{}
\maketitle

\begin{abstract}
We consider  energy-constrained infinite-dimensional quantum channels from a given system (satisfying a certain condition) to any other systems. We show that dealing with basic capacities of these channels we may assume (accepting arbitrarily small error $\varepsilon$) that all channels have the same finite-dimensional input space -- the subspace corresponding to the $m(\varepsilon)$ minimal eigenvalues of the input Hamiltonian.

We also show that for the class of energy-limited channels (mapping energy-bounded states to energy-bounded states) the above result is valid with substantially smaller dimension $m(\varepsilon)$.

The uniform finite-dimensional approximation allows to prove the uniform continuity of the basic capacities on the set of all quantum channels  with respect to the strong (pointwise) convergence topology. For all the capacities we obtain continuity bounds depending only on the input energy bound and the energy-constrained-diamond-norm distance between quantum channels (generating the strong convergence on the set of quantum channels).
\end{abstract}

\tableofcontents

\section{Introduction}

When we consider transmission of classical or  quantum  information over
infinite-dimensional quantum channels we have to impose energy constraints on states
used for encoding information to be consistent with the physical
implementation of the process  \cite{H-SCI,H-c-w-c,Wilde+}.

The energy constraint for a single channel $\,\Phi:A\rightarrow B\,$ is expressed  by the inequality
\begin{equation}\label{lc-}
\mathrm{Tr}H_A\rho \leq E,\quad \rho\in\S(\H_A),
\end{equation}%
where $H_A$ is the Hamiltonian of the input system $A$.

We will assume that $H_A$ is a positive operator having discrete spectrum $\{E^A_k\}_{k\geq0}$ of finite multiplicity such that $\,E^A_k\rightarrow+\infty$ as $k\rightarrow+\infty$.\footnote{This assumption holds for quantum systems used in applications, in particular, for a system of quantum oscillators.} In this case  $H_A$ determines the special family $\{\H^m_A\}_{m=1}^{+\infty}$ of  finite-dimensional subspaces of the input space $\H_A$, where $\H^m_A$ is the linear hull of the eigenvectors of $H_A$ corresponding to its $m$ minimal eigenvalues. The subspace $\H^m_A$ can be treated as the minimal energy $m$-dimensional subspace of $\H_A$. So, the states supported by $\H^m_A$ are more relevant to the constraint (\ref{lc-}) than states supported by other $m$-dimensional subspaces  of $\H_A$. At the same time, it is easy to show that all the states satisfying (\ref{lc-}) can be uniformly approximated by states in $\S(\H^m_A)$ satisfying (\ref{lc-}) for large $m$.

So, it is reasonable to ask what happens if we will use for encoding information only states in $\S(\H^m_A)$ satisfying (\ref{lc-}) for sufficiently large $m$ (for block encoding this means the use of the states in $\S([\H^m_A]^{\otimes n})$ satisfying (\ref{lc-}) with $H_A$ replaced by the Hamiltonian of $n$ copies of $A$ and $E$ replaced by $nE$).

It is clear that this additional restriction on the choice of codes-states (we will call it the \emph{$m$-restriction}) can not increase the ultimate rate of information transmission through a channel. On the other hand, the above observations  give a reason to conjecture that the loss of the information transmission rate caused by the $m$-restriction can be made arbitrarily small by increasing $m$.

For a \emph{fixed} channel $\Phi$ this conjecture can be easily verified for each of the basic capacities either by using operational definition of the capacity or by exploiting expressions of this capacity via entropic characteristics of a channel. In the paper we prove the channel-independent version of this assertion: the loss of each of the basic capacities caused by the $m$-restriction tends to zero as $m\rightarrow+\infty$ \emph{uniformly on the set of all channels} from the system $A$ to any other systems provided the Hamiltonian $H_A$ satisfies the condition\footnote{I would be grateful for any comments concerning physical sense of this condition.}
\begin{equation*}
  \lim_{\lambda\rightarrow+0}\left[\Tr e^{-\lambda H_A}\right]^{\lambda}=1,
\end{equation*}
which holds, in particular, for a system of quantum oscillators playing a central role in continuous variable quantum information theory \cite{H-SCI,W&Co}. \smallskip

We also show that the vanishing rate of the loss of the basic capacities caused by the $m$-restriction can be increased substantially by restricting attention to the class of quantum channels mapping energy-bounded states to energy-bounded states (called  energy-limited channels in \cite{W-EBN}).\smallskip

The uniform finite-dimensional approximation  allows to prove the uniform continuity of the basic capacities on the set of all quantum channels  with respect to the strong (pointwise) convergence topology.\footnote{I would be grateful for any comments about other applications of the uniform finite-dimensional approximation of energy-constrained channel capacities.}

\section{Preliminaries}

Let $\mathcal{H}$ be a separable infinite-dimensional Hilbert space,
$\mathfrak{B}(\mathcal{H})$ the algebra of all bounded operators  in $\mathcal{H}$ with the operator norm $\|\cdot\|$ and $\mathfrak{T}( \mathcal{H})$ the
Banach space of all trace-class
operators in $\mathcal{H}$ with the trace norm $\|\!\cdot\!\|_1$. Let
$\mathfrak{S}(\mathcal{H})$ be  the set of quantum states (positive operators
in $\mathfrak{T}(\mathcal{H})$ with unit trace) \cite{H-SCI,Wilde}.

Denote by $I_{\H}$ the unit operator in a Hilbert space
$\mathcal{H}$ and by $\id_{\mathcal{\H}}$ the identity
transformation of the Banach space $\mathfrak{T}(\mathcal{H})$.\smallskip

We will repeatedly use the inequality
\begin{equation}\label{b-ineq}
\|(I_{\H}-P)\,\rho\, P\|_1\leq \sqrt{\Tr(I_{\H}-P)\rho}
\end{equation}
valid for any state $\rho\in\S(\H)$ and any orthogonal projector $P\in\B(\H)$, which can be
easily proved via the operator Cauchy-Schwarz inequality (see the proof of Lemma 11.1 in \cite{H-SCI}).\smallskip

If quantum systems $A$ and $B$ are described by Hilbert spaces  $\H_A$ and $\H_B$ then the bipartite system $AB$ is described by the tensor product of these spaces, i.e. $\H_{AB}=\H_A\otimes\H_B$. A state in $\S(\H_{AB})$ is denoted $\rho_{AB}$, its marginal states $\Tr_B\rho_{AB}$ and $\Tr_A\rho_{AB}$ are denoted $\rho_{A}$ and $\rho_{B}$ respectively.\footnote{Here and in what follows $\Tr_X$ means $\Tr_{\H_X}$.} \smallskip

A \emph{quantum channel} $\,\Phi$ from a system $A$ to a system
$B$ is a completely positive trace preserving linear map from
$\mathfrak{T}(\mathcal{H}_A)$ into $\mathfrak{T}(\mathcal{H}_B)$ \cite{H-SCI,Wilde}.\smallskip

For any  quantum channel $\,\Phi:A\rightarrow B\,$ the Stinespring theorem implies existence of a Hilbert space
$\mathcal{H}_E$ and of an isometry
$V_{\Phi}:\mathcal{H}_A\rightarrow\mathcal{H}_B\otimes\mathcal{H}_E$ such
that
\begin{equation}\label{St-rep}
\Phi(\rho)=\mathrm{Tr}_{E}V_{\Phi}\rho V_{\Phi}^{*},\quad
\rho\in\mathfrak{T}(\mathcal{H}_A).
\end{equation}
The quantum  channel
\begin{equation}\label{c-channel}
\mathfrak{T}(\mathcal{H}_A)\ni
\rho\mapsto\widehat{\Phi}(\rho)=\mathrm{Tr}_{B}V_{\Phi}\rho
V_{\Phi}^{*}\in\mathfrak{T}(\mathcal{H}_E)
\end{equation}
is called \emph{complementary} to the channel $\Phi$
\cite[Ch.6]{H-SCI}.\smallskip

In finite dimensions (i.e. when $\dim\H_A$ and $\dim\H_B$ are finite) the distance  between quantum channels from $A$ to $B$ generated by the diamond norm
\begin{equation}\label{d-norm}
\|\Phi\|_{\diamond}\doteq \sup_{\rho\in\S(\H_{AR})}\|\Phi\otimes \id_R(\rho)\|_1
\end{equation}
of a Hermitian-preserving superoperator $\Phi:\T(\H_A)\rightarrow\T(\H_B)$ is widely  used \cite{Kit,Paul,Wilde}.
But this metric becomes singular in the case $\dim\H_A=\dim\H_B=+\infty$: there are infinite-dimensional channels with close physical parameters such that the diamond-norm distance between them equals to $2$ \cite{W-EBN}. In this case it is natural to use the distance between quantum channels generated by the energy-constrained diamond norm
\begin{equation}\label{ecd}
\|\Phi\|^E_{\diamond}\doteq \sup_{\rho\in\S(\H_{AR}), \Tr H_A\rho\leq E}\|\Phi\otimes \id_R(\rho)\|_1,\quad E>E^A_0,
\end{equation}
of a Hermitian-preserving superoperator $\Phi:\T(\H_A)\rightarrow\T(\H_B)$, where $H_A$ is the Hamiltonian of the input system $A$ and $E^A_0\doteq\inf\limits_{\|\varphi\|=1}\langle\varphi|H_A|\varphi\rangle$ \cite{SCT,W-EBN}.\smallskip

The \emph{von Neumann entropy} $H(\rho)=\mathrm{Tr}\eta(\rho)$ of a
state $\rho\in\mathfrak{S}(\mathcal{H})$, where $\eta(x)=-x\log x$ if $x>0$ and $\eta(0)=0$,
is a concave nonnegative lower semicontinuous function on the set $\mathfrak{S}(\mathcal{H})$ \cite{H-SCI,L-2,W}. The concavity of the von Neumann entropy is supplemented by the
inequality
\begin{equation}\label{w-k-ineq}
H(p\rho+(1-p)\sigma)\leq pH(\rho)+(1-p)H(\sigma)+h_2(p),
\end{equation}
where $h_2(p)=\eta(p)+\eta(1-p)$ is the binary entropy, valid for any
states  $\rho,\sigma\in\S(\H)$ and $p\in(0,1)$ \cite{H-SCI,Wilde}.
\smallskip

The \emph{quantum conditional entropy}
\begin{equation}\label{c-e-d}
H(A|B)_{\rho}=H(\rho_{AB})-H(\rho_B)
\end{equation}
of a bipartite state $\rho_{AB}$ with finite marginal entropies is essentially used in analysis of quantum systems \cite{H-SCI,Wilde}. It is concave and satisfies the following inequality
\begin{equation}\label{ce-ac}
H(A|B)_{p\rho+(1-p)\sigma}\leq p H(A|B)_{\rho}+(1-p)H(A|B)_{\sigma}+h_2(p)
\end{equation}
for any states $\rho,\sigma\in\S(\H_{AB})$ and $p\in(0,1)$. Inequality (\ref{ce-ac}) follows from concavity of the entropy and  inequality (\ref{w-k-ineq}).\smallskip

The \emph{quantum relative entropy} for two states $\rho$ and
$\sigma$ in $\mathfrak{S}(\mathcal{H})$ is defined as
$$
H(\rho\shs\|\shs\sigma)=\sum_i\langle
i|\,\rho\log\rho-\rho\log\sigma\,|i\rangle,
$$
where $\{|i\rangle\}$ is the orthonormal basis of
eigenvectors of the state $\rho$ and it is assumed that
$H(\rho\shs\|\shs\sigma)=+\infty$ if the support of $\rho\shs$ is not
contained in the support of $\sigma$ \cite{H-SCI,L-2,W}.\footnote{The support of a positive operator is the orthogonal complement of its kernel.}\smallskip

The \emph{quantum mutual information} of a state $\,\rho_{AB}\,$ of a
bipartite quantum system  is defined as
\begin{equation}\label{mi-d}
I(A\!:\!B)_{\rho}=H(\rho_{AB}\shs\Vert\shs\rho_{A}\otimes
\rho_{B})=H(\rho_{A})+H(\rho_{B})-H(\rho_{AB}),
\end{equation}
where the second expression  is valid if $\,H(\rho_{AB})\,$ is finite \cite{L-MI,Wilde}.\smallskip

Basic properties of the relative entropy show that $\,\rho\mapsto
I(A\!:\!B)_{\rho}\,$ is a lower semicontinuous function on the set
$\S(\H_{AB})$ taking values in $[0,+\infty]$. It is well known that
\begin{equation}\label{MI-UB}
I(A\!:\!B)_{\rho}\leq 2\min\left\{H(\rho_A),H(\rho_B)\right\}
\end{equation}
for any state $\rho_{AB}$ \cite{L-MI,Wilde}.\smallskip

By using the quantum mutual information the conditional entropy (\ref{c-e-d}) can be extended the set of all  bipartite states $\rho_{AB}$ with finite $H(\rho_{A})$ as follows
\begin{equation}\label{ext-ce}
H(A|B)_{\rho}=H(\rho_{A})-I(A\!:\!B)_{\rho}.
\end{equation}
This extension preserves all the basic properties of the conditional entropy (including concavity and inequality (\ref{ce-ac})) \cite{Kuz},\cite[Sect.5]{CMI}.
\smallskip

A finite or
countable collection $\{\rho_{i}\}$ of  states
with the corresponding probability distribution $\{p_{i}\}$ is conventionally called \emph{ensemble} and denoted $\{p_i,\rho_i\}$. The state $\bar{\rho}=\sum_{i} p_i\rho_i$ is called the \emph{average state} of this ensemble.
\smallskip

The Holevo quantity of an ensemble $\{p_i,\rho_i\}_{i=1}^m$ of $\,m\leq+\infty$ quantum states is defined as
$$
\chi\left(\{p_i,\rho_i\}_{i=1}^m\right)\doteq \sum_{i=1}^m p_i H(\rho_i\|\bar{\rho})=H(\bar{\rho})-\sum_{i=1}^m p_i H(\rho_i),
$$
where the second formula is valid if $H(\bar{\rho})<+\infty$. This quantity plays important role in analysis of information properties of quantum systems and channels \cite{H-SCI,Wilde}.\smallskip

Let $\H_A=\H$ and $\,\{|i\rangle\}_{i=1}^m$ be an orthonormal basis in a $m$-dimensional Hilbert space $\H_B$. Then
\begin{equation}\label{chi-rep}
\chi(\{p_i,\rho_i\}_{i=1}^m)=I(A\!:\!B)_{\hat{\rho}},\textrm{ where }\,\hat{\rho}_{AB}=\sum_{i=1}^m p_i\rho_i\otimes |i\rangle\langle i|.
\end{equation}

The \emph{quantum conditional mutual information (QCMI)} of a state $\rho_{ABC}$ of a
tripartite finite-dimensional system  is defined as
\begin{equation}\label{cmi-d}
    I(A\!:\!B|C)_{\rho}\doteq
    H(\rho_{AC})+H(\rho_{BC})-H(\rho_{ABC})-H(\rho_{C}).
\end{equation}
This quantity plays important role in quantum
information theory \cite{D&J,Wilde}, its nonnegativity is a basic result well known as \emph{strong subadditivity
of von Neumann entropy} \cite{Simon}. If system $C$ is trivial then (\ref{cmi-d}) coincides with (\ref{mi-d}).\smallskip

In infinite dimensions formula (\ref{cmi-d}) may contain the uncertainty
$"\infty-\infty"$. Nevertheless the
conditional mutual information can be defined for any state
$\rho_{ABC}$ by one of the equivalent expressions
\begin{equation}\label{cmi-e+}
\!I(A\!:\!B|C)_{\rho}=\sup_{P_A}\left[\shs I(A\!:\!BC)_{Q_A\rho
Q_A}-I(A\!:\!C)_{Q_A\rho Q_A}\shs\right],\; Q_A=P_A\otimes I_{BC},\!
\end{equation}
\begin{equation}\label{cmi-e++}
\!I(A\!:\!B|C)_{\rho}=\sup_{P_B}\left[\shs I(B\!:\!AC)_{Q_B\rho
Q_B}-I(B\!:\!C)_{Q_B\rho Q_B}\shs\right],\; Q_B=P_B\otimes I_{AC},\!
\end{equation}
where the suprema are over all finite rank projectors
$P_A\in\B(\H_A)$ and\break $P_B\in\B(\H_B)$ correspondingly and it is assumed that $I(X\!:\!Y)_{Q_X\rho
Q_X}=\lambda I(X\!:\!Y)_{\lambda^{-1} Q_X\rho
Q_X}$, where $\lambda=\Tr\shs Q_X\rho_{ABC}$ \cite{CMI}.\smallskip

Expressions (\ref{cmi-e+}) and
(\ref{cmi-e++}) define the same  lower semicontinuous function on the set
$\S(\H_{ABC})$ possessing all basic properties of the quantum conditional mutual
information valid in finite dimensions \cite[Th.2]{CMI}. In particular, the following relation (chain rule)
\begin{equation}\label{chain}
I(X\!:\!YZ|C)_{\rho}=I(X\!:\!Y|C)_{\rho}+I(X\!:\!Z|YC)_{\rho}
\end{equation}
holds for any state $\rho$ in $\S(\H_{XYZC})$ (with possible values $+\infty$ in both sides).
To prove (\ref{chain}) is suffices to note that it holds if the systems $X,Y,Z$ and $C$ are finite-dimensional and to apply Corollary 9 in \cite{CMI}.

We will use the upper bound
\begin{equation}\label{CMI-UB}
I(A\!:\!B|C)_{\rho}\leq 2\min\left\{H(\rho_A),H(\rho_B),H(\rho_{AC}),H(\rho_{BC})\right\}
\end{equation}
valid for any state $\rho_{ABC}$. It directly follows from upper bound (\ref{MI-UB}) and the expression
$I(X\!:\!Y|C)_{\rho}=I(X\!:\!YC)_{\rho}-I(X\!:\!C)_{\rho}$, $X,Y=A,B$, which is a partial case of (\ref{chain}). \smallskip

The quantum conditional  mutual information is not concave or convex but the following relation
\begin{equation}\label{F-c-b}
\begin{array}{cc}
\left|p
I(A\!:\!B|C)_{\rho}+(1-p)I(A\!:\!B|C)_{\sigma}-I(A\!:\!B|C)_{p\rho+(1-p)\sigma}\right|\leq h_2(p)
\end{array}
\end{equation}
holds for $p\in(0,1)$ and any states $\rho,\sigma\in\S(\H_{ABC})$ with finite QCMI.
If $\rho$ and $\sigma$ are states with finite marginal entropies then (\ref{F-c-b}) can be easily proved by noting that
\begin{equation*}
I(A\!:\!B|C)_{\rho}=H(A|C)_{\rho}-H(A|BC)_{\rho},
\end{equation*}
and by using  concavity of the conditional entropy and inequality
(\ref{ce-ac}). The validity of inequality (\ref{F-c-b}) for any states $\rho$ and $\sigma$  with finite QCMI is proved by approximation (using Theorem 2B in \cite{CMI}).\smallskip

Let $H_A$ be a positive operator in a Hilbert space $\H_A$ treated as a Hamiltonian of quantum system $A$.  Then $\,\Tr H_A\rho$ is the (mean) energy of a state $\rho\in\S(\H_A)$.\footnote{The value of $\,\Tr H_A\rho$ (finite or infinite) is defined as $\,\sup_n \Tr P_n H_A\rho$, where $P_n$ is the spectral projector of $H_A$ corresponding to the interval $[0,n]$.} So,
$$
\C_{H_A,E}=\{\rho\in\S(\H_A)\,|\,\Tr H_A\rho\leq E\},\quad E\geq E^A_0\doteq\inf\limits_{\|\varphi\|=1}\langle\varphi|H_A|\varphi\rangle,
$$
is a closed convex subset of $\S(\H_A)$ consisting of states with mean energy not exceeding $E$.\smallskip

It is well known that the von Neumann entropy is continuous on the set $\C_{H_A,E}$ for any $E\geq E^A_0$ if (and only if) the Hamiltonian  $H_A$ satisfies  the condition
\begin{equation}\label{H-cond}
  \Tr e^{-\lambda H_{A}}<+\infty\quad\textrm{for all}\;\,\lambda>0
\end{equation}
and that it achieves the maximal value on this set at the \emph{Gibbs state} $\gamma_A(E)=e^{-\lambda(E) H_A}/\Tr e^{-\lambda(E) H_A}$, where the parameter $\lambda(E)$ is determined by the equality $\Tr H_A e^{-\lambda(E) H_A}=E\Tr e^{-\lambda(E) H_A}$ \cite{W}.\smallskip

Condition (\ref{H-cond}) implies that $H_A$ is an unbounded operator having a discrete spectrum of finite multiplicity, i.e. it can be represented as follows
\begin{equation}\label{H-form}
H_A=\sum_{k=0}^{+\infty}E^A_k|\tau_k\rangle\langle \tau_k|,
\end{equation}
where $\{E^A_k\}$ is the nondecreasing sequence of eigenvalues of $H_A$ tending to $+\infty$ and $\{|\tau_k\rangle\}$ -- the corresponding basis of eigenvectors.\smallskip

In what follows we will use the function
\begin{equation*}
F_{H_A}(E)\doteq\sup_{\rho\in\C_{H_{\!A},E}}H(\rho)=H(\gamma_A(E)).
\end{equation*}
It is easy to show that $F_{H_A}$ is a strictly increasing concave function on $[E^A_0,+\infty)$ such that $F_{H_A}(E_0)=\log d_0$, where $d_0$ is the multiplicity of the eigenvalue $E^A_0$ \cite{CHI,W-CB}. \smallskip

In this paper we will use the  modification of the Alicki-Fannes-Winter method\footnote{This method is widely used in finite-dimensions for proving uniform continuity of functions on the set of quantum states \cite{A&F,W-CB}.} adapted for the set of states with bounded energy  \cite{AFM}. This modification makes it possible to prove uniform continuity of any locally almost affine function\footnote{This means that  $\,|f(p\rho+(1-p)\sigma)-p f(\rho)-(1-p)f(\sigma)|\leq r(p)=o(1)\,$ as $\,p\rightarrow+0$.} $\,f\,$ on the set
$$
\C^{\,\mathrm{ext}}_{H_A,E}\doteq\{\shs\rho\in\S(\H_{AB})\shs|\,\rho_A\in \C_{H_A,E}\shs\}\qquad (B\; \textrm{ is any given system})
$$
such that $|f(\rho_{AB})|\leq C H(\rho_A)$ for some $C\in\mathbb{R}_+ $ provided that
\begin{equation}\label{H-cond++}
  F_{H_A}(E)=o\shs(\sqrt{E})\quad\textrm{as}\quad E\rightarrow+\infty.
\end{equation}
By Lemma 1 in \cite{AFM} condition (\ref{H-cond++}) holds if and only if
\begin{equation}\label{H-cond+}
  \lim_{\lambda\rightarrow+0}\left[\Tr e^{-\lambda H_A}\right]^{\lambda}=1.
\end{equation}
Condition (\ref{H-cond+}) is  stronger than condition (\ref{H-cond}) (equivalent to  $\,F_{H_A}(E)=o\shs(E)\,$) but the difference between these conditions  is not too large. In terms of the sequence $\{E^A_k\}$ of eigenvalues of $H_A$
condition (\ref{H-cond}) means that $\lim_{k\rightarrow\infty}E^A_k/\log k=+\infty$, while (\ref{H-cond+}) is valid  if $\;\liminf_{k\rightarrow\infty} E^A_k/\log^q k>0\,$ for some $\,q>2$ \cite[Pr.1]{AFM}. \smallskip

It is essential that condition (\ref{H-cond+})  holds for the Hamiltonian of the multi-mode quantum oscillator playing central role in continuous variable quantum information theory \cite{H-SCI,W&Co}. \smallskip

If $A$ is the $\,\ell$-mode quantum oscillator with frequencies $\,\omega_1,...,\omega_{\ell}\,$ then
$$
F_{H_A}(E)=\max_{\{E_i\}}\sum_{i=1}^{\ell}g(E_i/\hbar\omega_i-1/2),\quad E\geq E_0\doteq\frac{1}{2}\sum_{i=1}^{\ell}\hbar\omega_i,\vspace{-5pt}
$$
where $\,g(x)=(x+1)\log(x+1)-x\log x\,$ and the maximum is over all $\ell\textup{-}$tuples $E_1$,...,$E_{\ell}$  such that  $\sum_{i=1}^{\ell}E_i=E$ and $E_i\geq\frac{1}{2}\hbar\omega_i$  \cite[Ch.12]{H-SCI}\cite{W-CB}. The exact value of
$F_{H_A}(E)$ can be calculated by applying the Lagrange multiplier method which leads to a transcendental equation. But following \cite{W-CB} one can obtain $\varepsilon$-sharp upper bound for $F_{H_A}(E)$ by using the inequality $\,g(x)\leq\log(x+1)+1\,$  valid for all $\,x>0$. It implies\vspace{-5pt}
$$
F_{H_A}(E)\leq \max_{\sum_{i=1}^{\ell}E_i=E}\sum_{i=1}^{\ell}\log(E_i/\hbar\omega_i+1/2)+\ell.
$$
By calculating this maximum via the Lagrange multiplier method  we obtain
\begin{equation}\label{q-osc}
F_{H_A}(E)\leq \widehat{F}_{\ell,\omega}(E)\doteq\ell\log \frac{E+E_0}{\ell E_*}+\ell,\quad E_*=\left[\prod_{i=1}^{\ell}\hbar\omega_i\right]^{1/\ell}.\vspace{-5pt}
\end{equation}
It is clear that the function $\widehat{F}_{\ell,\omega}$ satisfies condition (\ref{H-cond++}). So, it  can be used in the role of $F_{H_A}$ in all the results obtained by the modified Alicki-Fannes-Winter method (in particular, in the below Lemmas \ref{S-CMI-CB} and \ref{b-lemma}).

\smallskip

We will use the following simple lemma (see Corollary 12 in \cite{W-CB}). \smallskip
\begin{lemma}\label{GWL} \emph{If $f$ is a concave nonnegative function on $[0,+\infty)$ then for any positive $x< y$ and any $z\geq0$ the following inequality holds}
$$
xf(z/x)\leq yf(z/y).
$$
\end{lemma}


\section{Basic lemmas}

In the following two lemmas essentially used in the paper we will employ the  function $\,g(x)\!\doteq\!(1+x)h_2\!\left(\frac{x}{1+x}\right)=(x+1)\log(x+1)-x\log x,\,x>0$.\smallskip

By applying the modification of the  Alicki-Fannes-Winter method mentioned in Section 2 to the QCMI defined in (\ref{cmi-e+}),(\ref{cmi-e++}) we obtain the following

\smallskip

\begin{lemma}\label{S-CMI-CB} \emph{Let $\shs\rho$ and  $\shs\sigma$ be states in $\,\S(\H_{ABCD})$ s.t.  $\frac{1}{2}\|\shs\rho-\sigma\|_1\leq\varepsilon<\frac{1}{2}$.\\ Let $\,\H_*$ be a subspace of $\,\H_{AD}$ containing the supports of $\rho_{AD}$ and $\sigma_{AD}$. If $\,\Tr H_*\rho_{AD},\,\Tr H_*\sigma_{AD}\leq E<+\infty$ for some  positive operator $H_*$ in $\H_*$  satisfying condition (\ref{H-cond+}) then $I(A\!:\!B|C)_{\rho}$ and $I(A\!:\!B|C)_{\sigma}$ are finite and
\begin{equation}\label{S-CMI-CB++}
|I(A\!:\!B|C)_{\rho}-I(A\!:\!B|C)_{\sigma}|\leq 2\sqrt{2\varepsilon}
F_{H_{*}}\!\left(E/\varepsilon\right)+2g(\sqrt{2\varepsilon}),
\end{equation}
where $\,F_{H_*}(E)\doteq \sup\{H(\rho)\,|\,\supp\rho\subseteq\H_*,\,\Tr H_*\rho\leq E\}$.}
\smallskip

\emph{If  $\,\rho_{BC}=\sigma_{BC}\,$ then  (\ref{S-CMI-CB++}) holds with $\,2g(\sqrt{2\varepsilon})\,$  replaced by $\,g(\sqrt{2\varepsilon})$.}
\smallskip

\emph{If  $\,\rho\,$ and  $\,\sigma\,$ are pure states then  (\ref{S-CMI-CB++}) and its specification for the case $\,\rho_{BC}=\sigma_{BC}\,$ hold with $\,\varepsilon\,$  replaced by $\,\varepsilon^2/2$.}

\end{lemma}\medskip

Since condition (\ref{H-cond+}) implies that $F_{H_*}(E)=o\shs(\sqrt{E})$ as $E\rightarrow+\infty$, the right hand side of (\ref{S-CMI-CB++}) tends to zero as $\,\varepsilon\!\rightarrow\!0^{+}$.\smallskip

\emph{Proof.} We may consider $I(A\!:\!B|C)$ as a function on $\S(\H_{BC}\otimes\H_*)$. Continuity bound (\ref{S-CMI-CB++}) and its specification for pure states $\rho$ and  $\sigma$ can be directly obtained from Proposition 1 in \cite{AFM}
by using inequality (\ref{F-c-b}) and the inequalities
\begin{equation*}
0\leq I(A\!:\!B|C)_{\omega}\leq I(A\!:\!BC)_{\omega}\leq I(AD\!:\!BC)_{\omega}\leq 2H(\omega_{AD})
\end{equation*}
valid for any state $\omega$ in $\S(\H_{ABCD})$, which follow from the basic properties of QCMI and upper bound (\ref{MI-UB}). \smallskip

To prove the specification of (\ref{S-CMI-CB++}) for the case  $\,\rho_{BC}=\sigma_{BC}\,$ we have to repeat several steps from the proof of Theorem 1 in \cite{AFM}.

Assume first that $\rank\rho_{B}=\rank\sigma_{B}<+\infty$. Then
\begin{equation}\label{I-rep+}
I(A\!:\!B|C)_{\omega}=H(B|C)_{\omega}-H(B|AC)_{\omega},\quad \omega=\rho,\sigma,
\end{equation}
where $H(X|Y)$ is the extended conditional entropy defined in (\ref{ext-ce}).

Let $\hat{\rho}$ and $\hat{\sigma}$ be purifications of the states $\rho$ and $\sigma$ such that $\frac{1}{2}\|\hat{\rho}-\hat{\sigma}\|_1=\delta\doteq\sqrt{2\varepsilon}$ and $\,\hat{\tau}_{\pm}=\delta^{-1}[\shs\hat{\rho}-\hat{\sigma}\shs]_{\pm}$. Then
\begin{equation}\label{omega-star}
\frac{1}{1+\delta}\,\rho+\frac{\delta}{1+\delta}\,\tau_-=\omega_*=
\frac{1}{1+\delta}\,\sigma+\frac{\delta}{1+\delta}\,\tau_+,
\end{equation}
where $\tau_{\pm}=[\hat{\tau}_{\pm}]_{ABCD}$ (see \cite{AFM}). It is easy to see that
$\rank[\tau_{\pm}]_{B}<+\infty$. So, representation (\ref{I-rep+}) holds for $\omega=\tau_{\pm}$ as well.
Since  the assumption $\,\rho_{BC}=\sigma_{BC}\,$ and (\ref{omega-star}) imply
$\,[\tau_{+}]_{BC}=[\tau_{-}]_{BC}\,$, we obtain from (\ref{I-rep+}) that
\begin{equation}\label{dI-rep}
I(A\!:\!B|C)_{\omega_1}-I(A\!:\!B|C)_{\omega_2}=H(B|AC)_{\omega_2}-H(B|AC)_{\omega_1}
\end{equation}
for $(\omega_1,\omega_2)=(\rho,\sigma),(\tau_+,\tau_-)$.

By applying concavity of the conditional entropy and inequality (\ref{ce-ac}) to the
convex decompositions (\ref{omega-star}) of $\,\omega_*$ and taking (\ref{dI-rep}) into account we obtain
$$
\begin{array}{c}
(1-p)\left[I(A\!:\!B|C)_{\rho}-I(A\!:\!B|C)_{\sigma}\right]=(1-p)\left[H(B|AC)_{\sigma}-H(B|AC)_{\rho}\right]\\\\
\leq p
\left[H(B|AC)_{\tau_-}
-H(B|AC)_{\tau_+}\right]+\shs
h_2(p)\\\\\;= p
\left[I(A\!:\!B|C)_{\tau_+}
-I(A\!:\!B|C)_{\tau_-}\right]+\shs
h_2(p),
\end{array}
$$
where $p=\frac{\delta}{1+\delta}$. Similarly,
$$
(1-p)\left[I(A\!:\!B|C)_{\sigma}-I(A\!:\!B|C)_{\rho}\right]\leq p
\left[I(A\!:\!B|C)_{\tau_-}-
I(A\!:\!B|C)_{\tau_+}\right]+\shs h_2(p).
$$
Since $0\leq I(A\!:\!B|C)\leq I(AD\!:\!B|C)$, these inequalities show  that the left hand side of (\ref{S-CMI-CB++}) does not exceed
\begin{equation}\label{t-ub}
\delta\max\left\{I(AD\!:\!B|C)_{\tau_-}, I(AD\!:\!B|C)_{\tau_+}\right\}+\shs g(\delta).
\end{equation}
By the proof of Theorem 1 in \cite{AFM} the assumption $\,\Tr H_*\rho_{AD},\Tr H_*\sigma_{AD}\leq E$ implies  $\,\Tr H_*[\tau_{\pm}]_{AD}\leq E/\varepsilon$. So, by using (\ref{CMI-UB}) we obtain
$$
I(AD\!:\!B|C)_{\tau_{\pm}}\leq 2H([\tau_{\pm}]_{AD})\leq 2F_{H_{*}}\!\left(E/\varepsilon\right)
$$
and hence the quantity in (\ref{t-ub}) does not exceed the right hand side of (\ref{S-CMI-CB++}) with
$\,2g(\sqrt{2\varepsilon})\,$  replaced by $\,g(\sqrt{2\varepsilon})$.

Assume now that $\rho$ and $\sigma$ are arbitrary states such that $\,\rho_{BC}=\sigma_{BC}$. Let $\{P^n_B\}$ be a sequence of finite rank projectors in $\H_B$ strongly converging to the unit operator $I_B$ . Consider two sequences consisting of the states
$$
\rho^n=r_n^{-1}P^n_B\otimes I_{ADC}\, \rho P^n_B\otimes I_{ADC}\quad\textrm{and} \quad\sigma^n=r_n^{-1}P^n_B\otimes I_{ADC}\, \sigma P^n_B\otimes I_{ADC},
$$
where $r_n=\Tr P^n_B \rho_B=\Tr P^n_B \sigma_B$ (here and in what follows we assume that $n$ is sufficiently large).
It is easy to see that $r_n\rho^n_{AD}\leq  \rho_{AD}$ and $r_n\sigma^n_{AD}\leq \sigma_{AD}$ for all $n$. So, we have
$$
\Tr H_*\rho^n_{AD},\,\Tr H_*\sigma^n_{AD}\leq r_n^{-1}E.
$$
Take any sequence $\{\varepsilon_n\}$ tending to $\varepsilon$  such that $\frac{1}{2}\|\shs\rho^n-\sigma^n\|_1\leq\varepsilon_n<\frac{1}{2}$ for all $n$.
Since  $\,\rho^n_{BC}=\sigma^n_{BC}\,$ and  $\rank \rho^n_{B}=\rank \sigma^n_{B}<+\infty$, the above part of the proof implies that
\begin{equation}\label{t-cb}
|I(A\!:\!B|C)_{\rho^n}-I(A\!:\!B|C)_{\sigma^n}|\leq 2\sqrt{2\varepsilon_n}
F_{H_{*}}\!\left(E/(r_n\varepsilon_n)\right)+g(\sqrt{2\varepsilon_n}).
\end{equation}
By using the lower semicontinuity of the function  $\,\omega\rightarrow I(A\!:\!B|C)_{\omega}\,$ and its monotonicity under local operations (Th.2 in \cite{CMI}) it is easy to show that
$$
\lim_{n\rightarrow\infty}I(A\!:\!B|C)_{\omega^n}=I(A\!:\!B|C)_{\omega},\quad \omega=\rho, \sigma.
$$
So, passing to the limit in (\ref{t-cb}) implies (\ref{S-CMI-CB++}) with $\,2g(\sqrt{2\varepsilon})\,$  replaced by  $\,g(\sqrt{2\varepsilon})$.
\smallskip

If $\rho$ and $\sigma$ are pure states then we can take pure states $\hat{\rho}=\rho\otimes\varrho$ and  $\hat{\sigma}=\sigma\otimes\varsigma$ such that $\frac{1}{2}\|\hat{\rho}-\hat{\sigma}\|_1=\varepsilon$ and repeat the above arguments. $\square$ \smallskip

By using Lemma \ref{S-CMI-CB} and the Leung-Smith telescopic trick from \cite{L&S} one can prove the following lemma in which we will assume that
$H_A$ is the Hamiltonian of system $A$ having form (\ref{H-form}). We will use the function
$$
\bar{F}_{H_{A}}(E)=F_{H_{A}}(E+E_0^A),\quad\textrm{where}\quad F_{H_A}(E)\doteq\sup_{\Tr H_A\rho\leq E}H(\rho),
$$
and the notations $\,\bar{E}=E-E_0^A$, $\,\bar{E}_m^A=E_m^A-E_0^A\,$ for all $\,m>0$.
\smallskip
\begin{lemma}\label{b-lemma} \emph{Let $\,\Pi_m(\rho)=P_m\rho P_m+[\Tr(I_A-P_m)\rho]|\tau_0\rangle\langle\tau_0|$, where  $P_m$ is the projector on the subspace $\H^m_A$ corresponding to the minimal $\,m$ eigenvalues $\,E^A_0,..,E^A_{m-1}$ of $H_A$ and $\tau_0$ is any eigenvector corresponding to the eigenvalue $E^A_0$ . Let $\,\rho\shs$ be a state  in $\,\S(\H^{\otimes n}_{A}\otimes\H_{R})$ such that $\,\sum_{k=1}^n\Tr H_A\rho_{A_k}\leq nE$. If $H_A$ satisfies condition (\ref{H-cond+}) then
\begin{equation}\label{b-lemma+}
\left|I(B^n\!:\!R)_{\Phi^{\otimes n}\otimes\id_{R}(\rho)}-I(B^n\!:\!R)_{\Psi_m^{\otimes n}\otimes\id_{R}(\rho)}\right|\leq n f(E, m),
\end{equation}
for any channel $\,\Phi:A\rightarrow B$, where $\,\Psi_m=\Phi\circ\Pi_m$ and
\begin{equation}\label{f-def}
f(E, m)\doteq 4\sqrt[4]{\frac{\bar{E}}{\bar{E}^A_m}}\bar{F}_{H_{A}}\!\!\left(\frac{1}{2}\sqrt{\bar{E} \bar{E}^A_m}\right)+g\!\left(\!2\sqrt[4]{\frac{\bar{E}}{\bar{E}^A_m}}\right)+\frac{32\bar{E}}{\bar{E}^A_m}
\bar{F}_{H_{A}}\!\!\left(\!\frac{\bar{E}^A_m}{16}\!\right).
\end{equation}
is a quantity tending  to zero as $\,m\rightarrow+\infty\,$ for each $E>E^A_0$.}
\smallskip

\emph{If $\,\bar{E}< \bar{E}^A_m/16\,$ and  $\,\Tr H_A\rho_{A_k}\leq E\,$ for all $\,k=\overline{1,n}\,$ then the last term in (\ref{f-def}) can be removed. If $\,n=1$ and $\,s\doteq \bar{E}/\bar{E}^A_m+\sqrt{\displaystyle\bar{E}/\bar{E}^A_m}< 1/2\,$ then $f(E, m)$ in (\ref{b-lemma+}) can be replaced  by the quantity}
$$
2\sqrt{2s}\bar{F}_{H_{A}}\!\!\left(\!\frac{\bar{E}}{s}\right)+g\!\left(\!\sqrt{2s}\right).
$$

\emph{If $A$ is the $\ell$-mode quantum oscillator with  frequencies $\,\omega_1,...,\omega_{\ell}\,$ then the function $\,\bar{F}_{H_A}(E)$ in all the above formulas can be replaced by its upper bound
$\widehat{F}_{\ell,\omega}(E+E^A_0)$, where $\widehat{F}_{\ell,\omega}(E)$ is defined in (\ref{q-osc}). In this case the sequence $\{E^A_k\}_{k\geq0}$  consists of the numbers $\sum_{i=1}^{\ell}\hbar\omega_i(n_i-1/2), n_1,...,n_{\ell}\in \mathbb{N}$ arranged in the nondecreasing order.}
\end{lemma}\smallskip

\begin{remark}\label{b-lemma-r}
The below proof of Lemma \ref{b-lemma} shows that its assertion can be generalized by replacing the quantum mutual information $I(B^n\!:\!R)$ in (\ref{b-lemma+}) by the (extended) quantum conditional mutual information $I(B^n\!:\!R|C)$ defined by the equivalent expressions
(\ref{cmi-e+}) and (\ref{cmi-e++}).
\end{remark}\medskip

\emph{Proof.} The assumption of the lemma implies $\,H(\rho_{A_k})<+\infty\,$ for $\,k=\overline{1,n}$.\smallskip

Let $E$ be an environment for the channel $\Phi$, so that the Stinespring representations (\ref{St-rep}) holds with some isometry
$V_{\Phi}$ from $\H_A$ into $\H_{BE}$.

Following  the Leung-Smith telescopic method from \cite{L&S} consider the states
$$
\sigma_k=\Phi^{\otimes k}\otimes\Psi_m^{\otimes (n-k)}\otimes\id_{R}(\rho),\quad k=0,1,...,n.
$$
We have
\begin{equation}\label{tel}
\!\!\!\!\!\begin{array}{c}
\displaystyle \left|I(B^n\!:\!R)_{\sigma_n}\!-I(B^n\!:\!R)_{\sigma_0}\right|\displaystyle=
\left|\sum_{k=1}^n I(B^n\!:\!R)_{\sigma_k}\!-I(B^n\!:\!R)_{\sigma_{k-1}}\right|\\ \leq \displaystyle \sum_{k=1}^n \left|I(B^n\!:\!R)_{\sigma_k}\!-I(B^n\!:\!R)_{\sigma_{k-1}}\right|.
\end{array}\!\!\!
\end{equation}
By using the chain rule (\ref{chain}) we obtain for each $k$
\begin{equation}\label{tel+}
\!\!\begin{array}{ll}
I(B^n\!:\!R)_{\sigma_k}\!-I(B^n\!:\!R)_{\sigma_{k-1}}&\!\!\!\!\!= I(B_1...B_{k-1}B_{k+1}...B_n \!:\!R)_{\sigma_k}\\\\& \!\!\!\!\!+\,I(B_k\!:\!R|B_1...B_{k-1}B_{k+1}...B_n)_{\sigma_k}\\\\&\!\!\!\!\!-\,
I(B_1...B_{k-1}B_{k+1}...B_n \!:\!R)_{\sigma_{k-1}}\\\\&\!\!\!\!\!-\,I(B_k\!:\!R|B_1...B_{k-1}B_{k+1}...B_n)_{\sigma_{k-1}}\\\\&\!\!\!\!\!=
I(B_k\!:\!R|B_1...B_{k-1}B_{k+1}...B_n)_{\sigma_k}\\\\&\!\!\!\!\!-\,
I(B_k\!:\!R|B_1...B_{k-1}B_{k+1}...B_n)_{\sigma_{k-1}},\!\!\!
\end{array}
\end{equation}
where it was  used that $\Tr_{B_k}\sigma_k=\Tr_{B_k}\sigma_{k-1}$. Note that the finite entropy of the states $\,\rho_{A_1},...,\rho_{A_n}$, upper bound (\ref{CMI-UB}) and monotonicity of the QCMI under local channels guarantee finiteness of all the terms in (\ref{tel}) and (\ref{tel+}).

To estimate the last difference in (\ref{tel+}) consider  the states
\begin{equation*}
\hat{\sigma}_k=V_{\Phi}^{\otimes n}\otimes I_{R} \,\varrho_k\; [V_{\Phi}^{\otimes n}]^*\otimes I_{R}
\end{equation*}
in $\S(\H_{B^nE^nR})$, where $\varrho_k=\id_A^{\otimes k}\otimes\Pi_m^{\otimes (n-k)}\otimes \id_R(\rho)$, $k=0,1,2,...,n$.
The state $\hat{\sigma}_k$ is an extension of the state $\sigma_k$ for each $k$, i.e. $\Tr_{E^n}\hat{\sigma}_k=\sigma_k$. Note that
$[\varrho_k]_{A_j}=\rho_{A_j}$ for $j\leq k$ and $[\varrho_k]_{A_j}=\Pi_m(\rho_{A_j})$ for $j>k$. Hence
\begin{equation}\label{E-est}
  \Tr H_A[\varrho_k]_{A_j}\leq x_j\doteq\Tr H_A\rho_{A_j}\quad\textrm{ for all }k\textrm{ and }j.
\end{equation}

By using monotonicity of the trace norm under action of a channel and Lemmas \ref{trn},\ref{u-est} below we obtain
\begin{equation}\label{norm-est}
\!\!\!\begin{array}{c}
\|\hat{\sigma}_k-\hat{\sigma}_{k-1}\|_1=\|\varrho_k-\varrho_{k-1}\|_1\\\\=\left\|\id_A^{\otimes k}\otimes\Pi_m^{\otimes (n-k)}\otimes \id_R\left(\rho-\id_A^{\otimes (k-1)}\otimes\Pi_m\otimes\id_A^{\otimes (n-k)}\otimes \id_R(\rho)\right)\right\|_1\\\\\leq \|\rho-\id_A^{\otimes (k-1)}\otimes\Pi_m\otimes\id_A^{\otimes (n-k)}\otimes \id_R(\rho)\|_1\\\\\leq 2\Tr(I_A-P_m)\rho_{A_k}+2\sqrt{\Tr(I_A-P_m)\rho_{A_k}}\leq 2\varepsilon_k,
\end{array}\!\!
\end{equation}
where $\,\varepsilon_k\doteq2\sqrt{\displaystyle\bar{x}_k/\bar{E}^A_m}$, $\bar{x}_k=x_k-E^A_0$. \smallskip

Let $N_1$ be the set of all indexes $k$ for which $\bar{x}_k< \bar{E}^A_m/16$ and $N_2=\{1,..,n\}\setminus N_1$.\footnote{Similar splitting is used in the proof of Lemma 7 in \cite{W-EBN}.} Let $n_i=\sharp (N_i)$ and $X_i=\frac{1}{n_i}\sum_{k\in N_i}x_k$ and $\bar{X}_i=X_i-E_0^A$, $i=1,2$. It follows from (\ref{tel}) and (\ref{tel+}) that the left hand  side of (\ref{b-lemma+}) do not exceed $S_1+S_2$, where
$$
S_i=\sum_{k\in N_i}|I(B_k\!:\!R|Y_k)_{\sigma_k}-
I(B_k\!:\!R|Y_k)_{\sigma_{k-1}}|,\quad Y_k=B_1...B_{k-1}B_{k+1}...B_n.
$$

For each $k\in N_1$ we have $\varepsilon_k< 1/2$. So, by using (\ref{E-est}) and (\ref{norm-est}) and by noting that $\,\Tr_{B_k}\sigma_k=\Tr_{B_k}\sigma_{k-1}\,$ we obtain from Lemma \ref{S-CMI-CB} with $\H_*=V_{\Phi}\H_{A_k}\subseteq\H_{B_kE_k}$ and $H_*=V_{\Phi}H_{A}V_{\Phi}^*-E^A_0I_{\H_*}$ that
\begin{equation*}
\begin{array}{c}
\displaystyle|I(B_k\!:\!R|Y_k)_{\sigma_k}-
I(B_k\!:\!R|Y_k)_{\sigma_{k-1}}|\leq 2\sqrt{2\varepsilon_k}\bar{F}_{H_{A}}(\bar{x}_k/\varepsilon_k)+g(\sqrt{2\varepsilon_k})\\\\
=\displaystyle 4\sqrt[4]{\bar{x}_k/\bar{E}^A_m}\,\bar{F}_{H_{A}}\!\!\left(\frac{1}{2}\sqrt{\bar{x}_k\bar{E}^A_m}\right)+g\!\left(\!2\sqrt[4]{\bar{x}_k/\bar{E}^A_m}\right).
\end{array}
\end{equation*}
 Hence, by using the concavity\footnote{The concavity of the function $\sqrt[4]{x}\,\bar{F}_{H_A}(\sqrt{x})$ follows from the concavity and nonnegativity the function $\,\bar{F}_{H_A}(x)$. This can be shown by calculation of the second derivative.} of the functions $\sqrt[4]{x}\,\bar{F}_{H_A}(\sqrt{x})$, $\sqrt[4]{x}$ and $g(x)$ along with the monotonicity of $g(x)$ we obtain
\begin{equation}\label{s-1}
\begin{array}{c}
\displaystyle S_1\leq \sum_{k\in N_1} 4\sqrt[4]{\bar{x}_k/\bar{E}^A_m}\,\bar{F}_{H_{A}}\!\!\left(\frac{1}{2}\sqrt{\bar{x}_k\bar{E}_m}\right)+\sum_{k\in N_1}g\!\left(\!2\sqrt[4]{\bar{x}_k/\bar{E}^A_m}\right)\\
\;\leq n_1\displaystyle4\sqrt[4]{\bar{X}_1/\bar{E}^A_m}\,\bar{F}_{H_{A}}\!\!\left(\frac{1}{2}\sqrt{\bar{X}_1 \bar{E}^A_m}\right)+n_1g\!\left(\!2\sqrt[4]{\bar{X}_1/\bar{E}^A_m}\right).
\end{array}
\end{equation}

For each $k\in N_2$ the inequality $I(B_k\!:\!R|Y_k)\leq I(B_kE_k\!:\!R|Y_k)$ and upper bound (\ref{CMI-UB}) imply
\begin{equation*}
\begin{array}{c}
|I(B_k\!:\!R|Y_k)_{\sigma_k}-
I(B_k\!:\!R|Y_k)_{\sigma_{k-1}}|\leq 2\max\{H([\hat{\sigma}_k]_{B_kE_k}),H([\hat{\sigma}_{k-1}]_{B_kE_k})\}\\\\=2\max\{H([\varrho_k]_{A_k}),H([\varrho_{k-1}]_{A_{k}})\}\leq 2F_{H_{A}}(x_k),
\end{array}
\end{equation*}
where the last inequality follows from (\ref{E-est}). Since $(n-n_2)X_1+n_2X_2\leq nE$ and $X_1\geq E^A_0$, we have $X_2\leq n\bar{E}/n_2+E^A_0$. So, by using concavity and monotonicity of the function $\,F_{H_A}$ we obtain
\begin{equation}\label{s-2}
S_2\leq\sum_{k\in N_2} 2F_{H_{A}}(x_k)\leq 2n_2F_{H_{A}}(X_2)\leq 2n_2\bar{F}_{H_{A}}(n\bar{E}/n_2).
\end{equation}
It is easy to see that $\bar{X}_1\leq \bar{E}$. Since $\bar{x}_k> \bar{E}^A_m/16$ for all $k\in N_2$ and $(n-n_2)E^A_0+\sum_{k\in N_2}\bar{x}_k+n_2E^A_0\leq\sum_{k\in N_1}x_k+\sum_{k\in N_2}x_k\leq nE$, we have $n_2/n\leq 16\bar{E}/\bar{E}^A_m$. So, it follows from (\ref{s-1}),(\ref{s-2}) and Lemma \ref{GWL} that
$$
\frac{S_1+S_2}{n}\leq 4\sqrt[4]{\frac{\bar{E}}{\bar{E}^A_m}}\bar{F}_{H_{A}}\!\!\left(\frac{1}{2}\sqrt{\bar{E} \bar{E}^A_m}\right)+g\!\left(\!2\sqrt[4]{\frac{\bar{E}}{\bar{E}^A_m}}\right)+\frac{32\bar{E}}{\bar{E}^A_m}\bar{F}_{H_{A}}\!\!\left(\frac{\bar{E}^A_m}{16}\right).
$$

The vanishing of the quantity $f(E,m)$ as $\,m\rightarrow+\infty\,$ follows from Lemma 1 in \cite{AFM} stating
the equivalence of (\ref{H-cond++}) and (\ref{H-cond+}).\smallskip

The assertion concerning the case $\,\Tr H_A\rho_{A_k}\leq E\,$ for all $\,k=\overline{1,n}\,$ follow from the above proof, since in this case the set $N_2$ is empty. In the case $\,n=1$ one can directly apply Lemma \ref{S-CMI-CB} with trivial $C$, $\H_*=V_{\Phi}\H_{A}\subseteq\H_{BE}$ and  $H_*=V_{\Phi}H_{A}V_{\Phi}^*-E^A_0I_{\H_*}$  by using (\ref{norm-est}) with $k=1$. $\square$

\smallskip


\begin{lemma}\label{trn}\emph{ Let $\,\Pi:A\rightarrow A$ be the channel defined by the formula $\,\Pi(\rho)=P\rho P+[\Tr(I_A-P)\rho]\tau$, $\rho\in\T(\H_A)$, where $P$ is an orthogonal projector and $\,\tau$ is any state in $\,\S(\H_A)$. Then for arbitrary state $\omega\in\S(\H_{AB})$, where $B$ is any system, the following inequality holds}
$$
\|\omega-\Pi\otimes\id_B(\omega)\|_1\leq 2\Tr(I_A-P)\omega_A+2\sqrt{\Tr(I_A-P)\omega_A}.
$$
\end{lemma}\smallskip

\emph{Proof.} The required inequality is easily obtained from  inequality (\ref{b-ineq}).

\smallskip

\begin{lemma}\label{u-est}\emph{ Let $\,H_A$ be a positive operator in $\H_A$ having form (\ref{H-form}) and $P_m$ the projector on the subspace $\H^m_A$ corresponding to the minimal $\,m$ eigenvalues $\,E^A_0,..,E^A_{m-1}$ of $H_A$. Then for any state $\rho\in\S(\H_A)$ such that $\Tr H_A\rho\leq E$ the following inequality holds}
$$
\Tr(I_A-P_m)\rho\leq (E-E^A_0)/(E^A_m-E^A_0).
$$
\end{lemma}\smallskip

\emph{Proof.} Since $\Tr(I_A-P_m)\rho=1-\Tr P_m\rho$, the required inequality follows directly from the inequalities 
$$
E^A_0\Tr P_m\rho\leq\Tr P_m H_A\rho,\quad E^A_m\Tr(I_A-P_m)\rho\leq\Tr(I_A-P_m)H_A\rho.\; \;\square
$$

\section{Capacities of energy-constrained infinite-dimensional channels and their approximation}

In this section we show that dealing with basic capacities of energy constrained infinite-dimensional channels from a given system  to any other systems  we may consider (accepting arbitrarily small error $\varepsilon>0$) that all these channels have \emph{the same finite-dimensional input space} -- the subspace corresponding to the minimal eigenvalues of the input Hamiltonian. For each of the capacities the dimension of this subspace is explicitly determined by $\varepsilon$.

\subsection{Survey of basic capacities}

When we consider transmission of classical or quantum information over
infinite-dimensional quantum channels we have to impose constraints on states
used for encoding information. A typical physically motivated constraint is
the requirement of boundedness of states-codes average energy.
For a single channel this constraint is expressed by the inequality
\begin{equation}\label{lc}
\mathrm{Tr}H_A\rho \leq E,\quad \rho\in\S(\H_A),
\end{equation}%
where $H_A$ is  the Hamiltonian of the
input quantum system $A$, for $n$-copies of a channel it
can be written as follows
\begin{equation}\label{lc+}
\mathrm{Tr}H_{A^n}\rho \leq nE,\quad \rho\in\S(\H^{\otimes n}_{A}),
\end{equation}%
where $\,H_{A^n}=H_A\otimes I_A\otimes\ldots\otimes I_A+\ldots+I_A\otimes \ldots\otimes I_A\otimes H_A\, $ is the Hamiltonian of the system $A^n$ ($n$ copies of $A$) \cite{H-SCI,H-c-w-c,Wilde+}.\smallskip

We will assume  that the Hamiltonian $H_A$ satisfies condition (\ref{H-cond}).
\smallskip

The Holevo capacity of a
channel $\Phi:A\rightarrow B$ with the (input) energy constraint is defined as:
\begin{equation*}
C_{\chi}(\Phi ,H_{\!A},E)=\sup_{\mathrm{Tr}H_A\bar{\rho} \leq
E}\chi(\{p_i,\Phi(\rho_i)\}),
\end{equation*}%
where the supremum is over all input ensembles $\{p_i,\rho_i\}$ with the average energy $\,\sum_ip_i\Tr H_A\rho_i=\Tr H_A\bar{\rho}\;$ not exceeding $E$. This quantity determines the ultimate rate of transmission of classical information through the channel $\Phi$ by using nonentangled block encoding, for many  channels it coincides with the classical capacity  under the energy constraint \cite{GHP,H-SCI,H-c-w-c}.\smallskip

Operational definition of the classical capacity of energy-constrained  infinite-dimensional channels is presented in \cite{H-c-w-c}. By the Holevo-Schumacher-Westmore-land theorem adapted for constrained channels (\cite[Proposition 3]{H-c-w-c}) the classical capacity of
any  channel $\,\Phi:A\rightarrow B\,$  with constraint (\ref{lc+}) is given by
the  regularized expression
\begin{equation*}
C(\Phi,H_A,E)=\lim_{n\rightarrow +\infty }n^{-1}C_{\chi}(\Phi^{\otimes n},H_{A^n},nE).
\end{equation*}

The entanglement-assisted classical capacity of a quantum channel determines the ultimate rate of transmission of classical information when an entangled state between the input and the output of a channel is used as an
additional resource (see details in \cite{H-SCI,Wilde}). Operational definition of the entanglement-assisted classical capacity of energy-constrained  infinite-dimensional channels is presented in \cite{H-c-w-c}. By the most general version of the Bennett-Shor-Smolin-Thaplyal
theorem for energy-constrained  infinite-dimensional channels (\cite[Theorem 1]{H-Sh-3}) the classical
entanglement-assisted capacity of any  channel
$\,\Phi:A\rightarrow B\,$ with constraint (\ref{lc+}) determined by arbitrary positive operator $H_A$ is given by the expression
\begin{equation*}
C_{\mathrm{ea}}(\Phi,H_A,E)=\sup_{\mathrm{Tr}H_A\rho \leq
E}I(\Phi,\rho),
\end{equation*}
in which $\shs I(\Phi, \rho)\shs$ is the quantum mutual information of a channel $\Phi$ at a state $\rho$ defined as
\begin{equation*}
 I(\Phi,\rho)=I(B\!:\!R)_{\Phi\otimes\mathrm{Id}_{R}(\hat{\rho})},
\end{equation*}
where $\mathcal{H}_R\cong\mathcal{H}_A$ and $\hat{\rho}\shs$
is a pure state in $\S(\H_{AR})$ such that $\hat{\rho}_{A}=\rho$.\smallskip

Detailed analysis of the energy-constrained quantum and private capacities in the context of general-type infinite-dimensional channels\footnote{There are many papers devoted to analysis of these capacities for Gaussian channels, see \cite{H&W,Wolf} and the surveys in \cite{W&Co,Wilde+}.} has been made recently by Wilde and Qi in \cite{Wilde+}. The results in \cite{Wilde+} and \cite{Wilde-pc} give considerable reasons to conjecture validity of the following generalizations of
the Lloyd-Devetak-Shor theorem and of the Devetak theorem to constrained infinite-dimensional channels:
\begin{itemize}
  \item
the quantum capacity of any channel
$\,\Phi:A\rightarrow B\,$ with constraint (\ref{lc+})
is given by
the  regularized expression
\begin{equation*}
Q(\Phi,H_A,E)=\lim_{n\rightarrow +\infty }n^{-1}\bar{Q}(\Phi^{\otimes n},H_{A^n},nE),
\end{equation*}
where $\,\bar{Q}(\Phi,H_A,E)\,$ is the supremum of the coherent information\break $\,I_c(\Phi,\rho)\doteq I(\Phi,\rho)-H(\rho)\,$ on the set of all input states $\rho\in \S(\H_A)$ satisfying (\ref{lc}). \smallskip
  \item the private  capacity of any  channel
$\,\Phi:A\rightarrow B\,$ with  constraint (\ref{lc+})
is given by
the  regularized expression
\begin{equation*}
C_\mathrm{p}(\Phi,H_A,E)=\lim_{n\rightarrow +\infty }n^{-1}\bar{C}_\mathrm{p}(\Phi^{\otimes n},H_{A^n},nE),
\end{equation*}
where
\begin{equation}\label{PHC-def-be}
\bar{C}_\mathrm{p}(\Phi,H_A,E)=\sup_{\mathrm{Tr}H_A\bar{\rho}\leq E}\left[\chi(\{p_i,\Phi(\rho_i)\})-\chi(\{p_i,\widehat{\Phi}(\rho_i)\})\right]
\end{equation}
(the supremum is over all input ensembles $\{p_i,\rho_i\}$ with the average energy not exceeding $E$ and $\widehat{\Phi}$ is the complementary channel to the channel $\Phi$ defined in (\ref{c-channel})).
\end{itemize}
\medskip

\subsection{Uniform finite-dimensional approximation theorem.}

Assume that $H_A$ is an unbounded operator in $\H_A$ with dense domain having discrete spectrum of finite multiplicity, i.e. it can be represented as follows
$$
H_A=\sum_{k=0}^{+\infty}E^A_k|\tau_k\rangle\langle \tau_k|,
$$
where  $\{E^A_k\}$ is the nondecreasing sequence of eigenvalues of $H_A$ tending to $+\infty$ and $\{|\tau_k\rangle\}$ -- the corresponding basis of eigenvectors.
Denote by $\H^m_A$ the linear span of the vectors $|\tau_0\rangle,...,|\tau_{m-1}\rangle$, i.e. $\H^m_A$ is the subspace  corresponding to the minimal $m$  eigenvalues of $H_A$ (taking the multiplicity into account). Let $P_m$ be the projector onto $\H^m_A$.\smallskip

For a given channel $\Phi:A\rightarrow B$ denote by $\Phi_m$ the restriction of $\Phi$ to the Banach space $\T(\H^m_A)$ of all operators in
$\T(\H_A)$ supported by $\H^m_A$. The channel $\Phi_m$ can be called the \emph{subchannel of $\,\Phi$ corresponding to the subspace $\H_A^m$}.
Since $H^m_A=P_m H_A$ is a positive (bounded) operator in $\B(\H_A^m)$, we may consider the capacities
$$
C_*(\Phi_m ,H^m_{\!A},E),\quad C_*=C_{\chi}, C, C_{\mathrm{ea}}, Q, C_{\mathrm{p}}.
$$
These capacities can be treated as the corresponding capacities of $\Phi$ obtained by block encoding used only states supported by the tensor powers of the $m$-dimensional subspace $\H^m_A$. We will call they \emph{$m$-restricted} capacities and will use the notations
$\;C_*^m(\Phi ,H_{\!A},E)\doteq C_*(\Phi_m ,H^m_{\!A},E)$, $C_*=C_{\chi}, C, C_{\mathrm{ea}}, Q, C_{\mathrm{p}}$.\smallskip

The following theorem states that any $m$-restricted capacity $C_*^m(\Phi ,H_{\!A},E)$ tends to the
 corresponding capacity $C_*(\Phi ,H_{\!A},E)$ as $m\rightarrow+\infty$ \emph{uniformly on the set of all channels} from a given system $A$ to any other systems
and gives explicit estimates for the rate of this convergence. In this theorem we use the function
$$
\bar{F}_{H_{A}}(E)=F_{H_{A}}(E+E_0^A),\quad\textrm{where}\quad F_{H_A}(E)\doteq\sup_{\Tr H_A\rho\leq E}H(\rho),
$$
and the notations $\,\bar{E}=E-E_0^A$, $\,\bar{E}_m^A=E_m^A-E_0^A\,$ for all $\,m>0$.
\smallskip\smallskip

\begin{theorem}\label{UFA} \emph{Let $C_*$ be one of the capacities $C_{\chi}, C, C_{\mathrm{ea}}, Q$ and $C_{\mathrm{p}}$. If the Hamiltonian $H_A$ satisfies condition (\ref{H-cond+}) and $E\geq E^A_0$ then for any $\,\varepsilon>0$ there exists natural number $\,m_{C_*}(\varepsilon)$  such that
\begin{equation}\label{C-approx}
|C_*(\Phi, H_{\!A},E)-C_*^m(\Phi, H_{\!A},E)|\leq\varepsilon\qquad \forall m\geq m_{C_*}(\varepsilon)
\end{equation}
for arbitrary  channel $\,\Phi$ from the system $A$ to any system $B$.}\smallskip

\emph{The above $\,m_{C_*}(\varepsilon)$ is the minimal natural number such that  $f_{C_*}(E,m)\leq\varepsilon$ and $\bar{E}^A_m\geq 16\bar{E}$, where}\footnote{$\,g(x)\!\doteq\!(1+x)h_2\!\left(\frac{x}{1+x}\right)=(x+1)\log(x+1)-x\log x$.}
$$
f_{C_{\chi}}(E,m)=2\sqrt{2s}\shs\bar{F}_{H_{A}}\!\!\left(\!\frac{\bar{E}}{s}\right)+g\!\left(\!\sqrt{2s}\right)\!,\quad s=\frac{\bar{E}}{\bar{E}^A_m}+\sqrt{\frac{\bar{E}}{\bar{E}^A_m}},
$$
$$
f_{C}(E,m)=4\sqrt[4]{\frac{\bar{E}}{\bar{E}^A_m}}\shs\bar{F}_{H_{A}}\!\!\left(\frac{1}{2}\sqrt{\bar{E} \bar{E}^A_m}\right)+g\!\left(\!2\sqrt[4]{\frac{\bar{E}}{\bar{E}^A_m}}\right)\!,
$$
$$
f_{C_{\mathrm{ea}}}(E,m)=2s\bar{F}_{H_{A}}\!\!\left(\!\frac{2\bar{E}}{s^2}\right)+2g(s),\quad s=\frac{\bar{E}}{\bar{E}^A_m}+\sqrt{\frac{\bar{E}}{\bar{E}^A_m}},
$$
$$
f_Q(E,m)=f_{C}(E,m)+\frac{32\bar{E}}{\bar{E}^A_m}\shs\bar{F}_{H_{A}}\!\!\left(\!\frac{\bar{E}^A_m}{16}\!\right)\!,\quad f_{C_{\mathrm{p}}}(E,m)=2f_Q(E,m).
$$

\emph{If $A$ is the $\ell$-mode quantum oscillator with frequencies $\,\omega_1,...,\omega_{\ell}\,$ then the function $\,\bar{F}_{H_A}$ in all the above formulas can be replaced by its upper bound
$\widehat{F}_{\ell,\omega}(E+E^A_0)$, where $\widehat{F}_{\ell,\omega}(E)$ is defined in (\ref{q-osc}). In this case the sequence $\{E^A_k\}_{k\geq0}$ consists of the numbers $\sum_{i=1}^{\ell}\hbar\omega_i(n_i-1/2),\, n_1,...,n_{\ell}\in \mathbb{N}$ arranged in the nondecreasing order.}
\end{theorem}\smallskip

\smallskip

\begin{remark}\label{UFA-r}
The existence of solutions of the inequalities $\,f_{C_*}(E,m)\leq\varepsilon,\,$ $\,C_*=C_{\chi},...,C_{\mathrm{p}},\,$ for any $\,\varepsilon>0\,$ is guaranteed by condition (\ref{H-cond+}), since it implies that $\,\bar{F}_{H_A}(E)=o(\sqrt{E})\,$ as $\,E\rightarrow+\infty\,$ by Lemma 1 in \cite{AFM}.
\end{remark}

\smallskip

The number $\,m_{C_*}(\varepsilon)$ will be called \emph{$\varepsilon$-sufficient input dimension} for $C_*$.

\medskip

\emph{Proof.} Let $P_m=\sum_{k=0}^{m-1}|\tau_k\rangle\langle \tau_k|$ be the projector on the subspace $\H^m_A$ and
 $\Pi_m:A\rightarrow A$  the channel introduced in Lemma \ref{b-lemma}.

\textbf{$C_*=C_{\chi}$}. If $\{p_i,\rho_i\}$ is an ensemble of input states such that $\Tr H_A\bar{\rho}\leq E$ then
the ensemble $\{p_i,\rho^m_i\}$, where $\rho^m_i=\Pi_m(\rho_i)$ for all $i$, satisfies the same condition for all $m$. So, the last assertion of Lemma \ref{b-lemma} and representation (\ref{chi-rep}) show that
$$
\left|\chi(\{p_i,\Phi(\rho_i)\})-\chi(\{p_i,\Phi(\rho^m_i)\})\right|\leq f_{C_{\chi}}(E,m).
$$
This implies the assertion of the theorem for $C_*=C_{\chi}$, since all the states $\rho^m_i$ are supported by the subspace $\H_A^m$. \smallskip

\textbf{$C_*=C$}. Note that
$$
C_{\chi}(\Phi^{\otimes n},H_{A^n},nE)=\sup\chi(\{p_i,\Phi^{\otimes n}(\rho_i)\}),
$$
where the supremum is over all ensembles  $\{p_i,\rho_i\}$ of states in $\S(\H^{\otimes n}_A)$ with the average state $\bar{\rho}$ such that $\Tr H_A\bar{\rho}_{A_j}\leq E$ for all $j=\overline{1,n}$. This can be easily shown by using the symmetry arguments and the following well known property of the Holevo quantity:
$$
\frac{1}{n}\sum_{j=1}^n\chi\left(\{q^j_i,\sigma^j_i\}_i\right)\leq \chi\left(\left\{\frac{q^j_i}{n},\sigma^j_i\right\}_{ij}\right)
$$
for any collection $\,\{q^1_i,\sigma^1_i\},...,\{q^n_i,\sigma^n_i\}\,$ of discrete ensembles. \smallskip

If $\{p_i,\rho_i\}$ is an ensemble of states in $\S(\H^{\otimes n}_A)$ satisfying the above condition then the ensemble
$\{p_i,\rho^m_i\}$, where $\rho^m_i=\Pi_m^{\otimes n}(\rho_i)$ for all $i$, satisfies the same condition for all $m$. So, the last assertion of Lemma \ref{b-lemma} and representation (\ref{chi-rep}) show that
$$
\left|\chi(\{p_i,\Phi^{\otimes n}(\rho_i)\})-\chi(\{p_i,\Phi^{\otimes n}(\rho^m_i)\})\right|\leq f_{C}(E,m).
$$
This implies the assertion of the theorem for $C_*=C$, since all the states $\rho^m_i$ are supported by the subspace $[\H_A^m]^{\otimes n}$. \smallskip

\textbf{$C_*=C_{\mathrm{ea}}$}. Let $\rho$ be any state in $\S(\H_A)$ such that $\Tr H_A\rho \leq
E$ and $\hat{\rho}$  its purification in $\S(\H_{AR})$. Then  $\rho_m\doteq (1-r_m)^{-1}P_m\rho P_m$, $r_m=1-\Tr P_m\rho$, is a state in $\S(\H_A)$ satisfying the same condition for all $m$ such that $E^A_m>E$ and $\hat{\rho}_m\doteq (1-r_m)^{-1}P_m\otimes I_R \,\hat{\rho}\, P_m\otimes I_R$ is a purification of this state. It follows from inequality (\ref{b-ineq}) that
$$
\!\|\hat{\rho}-\hat{\rho}_m\|_1\leq\|\hat{\rho}-P_m\otimes I_R\,\hat{\rho}\, P_m\otimes I_R\|_1+\|P_m\otimes I_R \,\hat{\rho}\, P_m\otimes I_R-\hat{\rho}_m\|_1\leq 2r_m+2\sqrt{r_m}.
$$
By Lemma \ref{u-est} the condition  $\Tr H_A\rho \leq
E$ implies $r_m\leq \bar{E}/\bar{E}^A_m\leq1/16$. So, by using the Stinespring representation (\ref{St-rep}) and the last assertion of Lemma \ref{S-CMI-CB} with trivial $C$, $\H_*=V_{\Phi}\H_{A}\subseteq\H_{BE}$ and  $H_*=V_{\Phi}H_{A}V_{\Phi}^*-E^A_0I_{\H_*}$  one can show that
$$
\left|I(B\!:\!R)_{\Phi\otimes\mathrm{Id}_{R}(\hat{\rho})}-I(B\!:\!R)_{\Phi\otimes\mathrm{Id}_{R}(\hat{\rho}_m)}\right|\leq 2s\bar{F}_{H_A}(2\bar{E}/s^2)+2g(s).
$$
This implies the assertion of the theorem for $C_*=C_{\mathrm{ea}}$, since the state $\rho_m$ is supported by the subspace $\H_A^m$. \smallskip

\textbf{$C_*=Q$}. Let $\Psi_m=\Phi\circ\Pi_m$, $\rho$ be any state in $\S(\H_A^{\otimes n})$ such that $\sum_{k=1}^n \Tr H_A\rho_{A_k}\leq nE$  and $\hat{\rho}$  its purification in $\S(\H_{A^nR})$. Then Lemma \ref{b-lemma} implies
$$
\left|I_c(\Phi^{\otimes n},\rho)-I_c(\Psi_m^{\otimes n},\rho)\right|=\left|I(B^n\!:\!R)_{\Phi^{\otimes n}\otimes\id_{R}(\hat{\rho})}-I(B^n\!:\!R)_{\Psi_m^{\otimes n}\otimes\id_{R}(\hat{\rho})}\right|\leq f_{Q}(E,m)
$$
This implies the assertion of the theorem for $C_*=Q$, since the operational definition of the quantum
capacity with the energy constraint (see Section III in \cite{Wilde+}) and the implication
\begin{equation}\label{p-imp}
\Tr H_{A^n}\rho\leq nE\quad \Rightarrow\quad \Tr H_{A^n}\Pi^{\otimes n}_m(\rho)\leq nE
\end{equation}
valid for any state $\rho\in\S(\H^{\otimes n}_{A})$ and $E\geq E^A_0$ show that
$$
Q(\Psi_m, H_{\!A},E)\leq Q_m(\Phi, H_{\!A},E)\leq Q(\Phi, H_{\!A},E).
$$

\textbf{$C_*=C_{\mathrm{p}}$}. If $\{p_i,\rho_i\}$ is an ensemble of states in $\S(\H^{\otimes n}_A)$ such that $\,\sum_{k=1}^n\Tr H_A \bar{\rho}_{A_k}\leq nE\,$  then the ensemble $\{p_i,\rho^m_i\}$, where $\rho^m_i=\Pi_m^{\otimes n}(\rho_i)$ for all $i$, satisfies the same condition for all $m$. So, Lemma \ref{b-lemma} and representation (\ref{chi-rep}) show that
$$
\left|\chi(\{p_i,\Phi^{\otimes n}(\rho_i)\})-\chi(\{p_i,\Phi^{\otimes n}(\rho^m_i)\})\right|\leq f_{C_{\mathrm{p}}}(E,m)/2.
$$
and
$$
\left|\chi(\{p_i,\widehat{\Phi}^{\otimes n}(\rho_i)\})-\chi(\{p_i,\widehat{\Phi}^{\otimes n}(\rho^m_i)\})\right|\leq f_{C_{\mathrm{p}}}(E,m)/2.
$$
This implies the assertion of the theorem for $C_*=C_{\mathrm{p}}$, since all the states $\rho^m_i$ are supported by the subspace $[\H_A^m]^{\otimes n}$. $\square$ \smallskip

Unfortunately, the values of $m_{C_*}(\varepsilon)$ given by Theorem \ref{UFA} for real physical systems are extremely high.

\begin{example}\label{exam}
Let $A$ be the one-mode quantum oscillator with the frequency $\omega$. In this case the Hamiltonian $H_A$ has the spectrum $\{E^A_k=(k+1/2)\hbar\omega\}_{k\geq 0}$ and $F_{H_A}(E)=g(E/\hbar\omega-1/2)$ \cite[Ch.12]{H-SCI}. The results of numerical calculations of $\,m_{C_*}(\varepsilon)$ for different values of the input energy bound $E$ are presented in the following tables corresponding to two values of the relative error $\varepsilon/F_{H_A}(E)$ equal respectively to $0.1$ and $0.01$.\footnote{The capacities  $C_*(\Phi, H_{\!A},E)$, $C_*=C_{\chi}, C, Q, C_{\mathrm{p}}$, take values in $[\shs0,F_{H_A}(E)\shs]$, the capacity $C_{\mathrm{ea}}(\Phi, H_{\!A},E)$ takes values in $[\shs0,2F_{H_A}(E)\shs]$.}\medskip

 Table 1. The approximate values of $m_{C_*}(\varepsilon)$ for $\varepsilon=0.1F_{H_A}(E)$.

\begin{tabular}{|c|c|c|c|c|c|}
  \hline
  $\;E/\hbar\omega\;$ & $m_{C_{\chi}}(\varepsilon)$ & $m_C(\varepsilon)$ & $m_{C_{\mathrm{ea}}}(\varepsilon)$ & $m_Q(\varepsilon)$ & $m_{C_{p}}(\varepsilon)$\\ \hline
  3 & $5.0\cdot10^{9}$ & $2.0\cdot10^{10}$ & $8.6\cdot10^{4}$ & $2.0\cdot10^{10}$ & $5.2\cdot10^{11}$\\ \hline
  10 & $3.2\cdot10^{9}$ & $1.3\cdot10^{10}$ & $1.3\cdot10^{5}$ & $1.3\cdot10^{10}$ & $3.4\cdot10^{11}$\\ \hline
  100 & $5.5\cdot10^{9}$ & $2.2\cdot10^{10}$ & $5.3\cdot10^{5}$ & $2.2\cdot10^{10}$ & $5.5\cdot10^{11}$\\
  \hline
\end{tabular}
\medskip

 Table 2. The approximate values of $m_{C_*}(\varepsilon)$ for $\varepsilon=0.01F_{H_A}(E)$.

\begin{tabular}{|c|c|c|c|c|c|}
  \hline
  $\;E/\hbar\omega\;$ & $m_{C_{\chi}}(\varepsilon)$ & $m_C(\varepsilon)$ & $m_{C_{\mathrm{ea}}}(\varepsilon)$ & $m_Q(\varepsilon)$ & $m_{C_{p}}(\varepsilon)$\\ \hline
  3 & $2.1\cdot10^{14}$ & $8.2\cdot10^{14}$ & $1.7\cdot10^{7}$ & $8.2\cdot10^{14}$ & $1.8\cdot10^{16}$\\ \hline
  10 & $1.3\cdot10^{14}$ & $5.3\cdot10^{14}$ & $2.6\cdot10^{7}$ & $5.3\cdot10^{14}$ & $1.7\cdot10^{16}$\\ \hline
  100 & $2.0\cdot10^{14}$ & $8.1\cdot10^{14}$ & $1.0\cdot10^{8}$ & $8.1\cdot10^{14}$ & $1.8\cdot10^{16}$\\
  \hline
\end{tabular}
\medskip

We see that the values of the $\varepsilon$-sufficient input dimension $m_{C_*}(\varepsilon)$ are extremely high for all the capacities excepting $C_{\mathrm{ea}}$. It is clear that this is explained by inaccuracy of the used estimates rather than physical reasons. In a sense, this is a cost of the universality of Theorem \ref{UFA} in which the class of \emph{all channels} from a given system $A$ to arbitrary systems $B$ are considered. In the next subsection we show that estimates of the $\varepsilon$-sufficient input dimension can be decreased substantially by restricting the class of channels $\Phi$ for which the validity of (\ref{C-approx}) is required.

\end{example}

\subsection{Specifications for energy-limited channels}

Theorem \ref{UFA} gives estimates of the $\varepsilon$-sufficient input dimensions for all quantum channels from a given system $A$ to arbitrary system $B$, which do not depend on system $B$ at all. Unfortunately, for a real quantum system (quantum oscillator) in the role of $A$ these estimates are extremely hight (see Example 1 and the comments below). In this section we show that estimates of the $\varepsilon$-sufficient input dimensions can be decreased substantially by imposing constraints on the class of quantum channels used for communications.

Assume that $B$ is a quantum system with the Hamiltonian $H_B$ satisfying condition (\ref{H-cond}) while $A$ is any quantum system with the Hamiltonian $H_A$ having form (\ref{H-form}). Consider quantum channels $\Phi$ from $A$ to $B$ such that
\begin{equation}\label{elc}
\Tr H_B\Phi(\rho)\leq \alpha \Tr H_A\rho +E_c\quad\textrm{for any}\; \rho\in\S(\H_A),
\end{equation}
where $\alpha$ and $E_c$ are nonnegative parameters. Such channels are called \emph{energy-limited} in \cite{W-EBN}, where it is mentioned that
any quantum channel mapping energy-bounded states to energy-bounded states satisfies (\ref{elc}) with some $\alpha$ and $E_c$.

Let $\widehat{F}_{H_B}$ be any upper bound for the function
\begin{equation*}
F_{H_B}(E)\doteq\sup_{\Tr H_B\rho\leq E}H(\rho)=H(\gamma_B(E)),\quad E\geq E^B_0\doteq\inf\limits_{\|\varphi\|=1}\langle\varphi|H_B|\varphi\rangle,
\end{equation*}
defined on $[0,+\infty)$ such that
\begin{equation}\label{F-prop-1B}
\widehat{F}_{H_B}(E)> 0,\quad \widehat{F}_{H_B}^{\shs\prime}(E)>0,\quad \widehat{F}_{H_B}^{\shs\prime\prime}(E)< 0\quad\textrm{for all }\; E>0.
\end{equation}
and
\begin{equation}\label{F-prop-2B}
\widehat{F}_{H_B}(E)=o\shs(E)\quad\textrm{as}\quad E\rightarrow+\infty.
\end{equation}
Since $H_B$ satisfies condition (\ref{H-cond}), one can use the function $E\mapsto F_{H_B}(E+E^B_0)$ in the role of $\widehat{F}_{H_B}$ \cite{CHI}.
If $B$ is the $\ell$-mode quantum oscillator with the frequencies  $\,\omega_i\,$ (\cite[Ch.12]{H-SCI}) then the function
$\widehat{F}_{\ell,\omega}$ defined in (\ref{q-osc}) also satisfies the above requirements for $\widehat{F}_{H_B}$.
\smallskip

Denote by $\F_{\alpha,E_c}(A,B)$ the class of all quantum channels from $A$ to $B$ satisfying (\ref{elc}). The following theorem is a version (specification) of Theorem \ref{UFA} for energy-limited channels and all the basic capacities excepting $C_{\mathrm{p}}$.
 \smallskip

\begin{theorem}\label{UFA-el} \emph{Let $C_*$ be one of the capacities $C_{\chi}, C, C_{\mathrm{ea}}$ and $Q$. If the Hamiltonian $H_B$ satisfies condition (\ref{H-cond}) and $E\geq E^A_0$ then for any $\alpha>0$, $E_c\geq0$ and $\,\varepsilon>0$ there exists natural number $\,m_{C_*}(\varepsilon|\,\alpha,E_c)$  such that
$$
|C_*(\Phi, H_{\!A},E)-C_*^m(\Phi, H_{\!A},E)|\leq\varepsilon\qquad \forall m\geq m_{C_*}(\varepsilon|\,\alpha,E_c)
$$
for arbitrary  channel $\,\Phi$ from the class $\,\F_{\alpha,E_c}(A,B)$.}\smallskip

\emph{If $\,C_*=C_{\chi}, C, C_{\mathrm{ea}}$ then $\,m_{C_*}(\varepsilon|\,\alpha,E_c)$ is the minimal natural number such that $f^{\alpha,E_c}_{C_*}(E,m|\,t)\leq\varepsilon$ for some $\,t\in(0,\frac{1}{2}]$, where
$$
f^{\alpha,E_c}_{C_{\chi}}(E,m|\,t)=(2t+s_m(t))\widehat{F}_{H_{B}}\!\left(\!\frac{\alpha E+E_c}{t}\right)+2g\!\left(s_m(t)\right)+2h_2(t),
$$
$$
f^{\alpha,E_c}_{C}(E,m|\,t)=f^{\alpha,E_c}_{C_{\mathrm{ea}}}(E,m|\,t)=(4t+2s_m(t))\widehat{F}_{H_{B}}\!\left(\!\frac{\alpha E+E_c}{t}\right)+2g\!\left(s_m(t)\right)+4h_2(t),
$$
where  $s_m(t)=\frac{\bar{E}/\bar{E}^A_m+\sqrt{\bar{E}/\bar{E}^A_m}+t/2}{1-t}$, $\bar{E}=E-E^A_0$, $\bar{E}^A_m=E^A_m-E^A_0$.}\medskip

\emph{The above $\,m_{Q}(\varepsilon|\,\alpha,E_c)$ is the minimal natural number s.t. $f^{\alpha,E_c}_{Q}(E,m|\,p,t)\leq\varepsilon$ for some $p>1$ and $\,t\in(0,\frac{1}{2}]$, where
$$
f^{\alpha,E_c}_{Q}(E,m|\,p,t)=(4t+2s_m(t))\widehat{F}_{H_{B}}\!\left(\!\frac{E_p}{t}\right)+2g\!\left(s_m(t)\right)+4h_2(t)+
\frac{2}{p}\widehat{F}_{H_{B}}(E_p),
$$
where $E_p=\alpha pE+E_c$.}
\end{theorem}\smallskip

\begin{remark}\label{UFA-el-r}
The existence of solutions of the inequalities determining $\,m_{C_*}(\varepsilon|\,\alpha,E_c)$, $\,C_*=C_{\chi},...,Q,\,$ for any $\,\varepsilon>0\,$ is guaranteed by condition (\ref{F-prop-2B}).
\end{remark}

\emph{Proof.} Let $\Psi_m=\Phi\circ\Pi_m$, where $\,\Pi_m:A\rightarrow A$ is  the channel defined in Lemma \ref{b-lemma}.  By Lemmas \ref{trn},\ref{u-est} and definition (\ref{ecd}) of the energy-constrained diamond norm we have
\begin{equation}\label{ecd-up}
\begin{array}{c}
\frac{1}{2}\|\Phi-\Psi_m\|_{\diamond}^E\leq\frac{1}{2}\|\id_A-\Pi_m\|_{\diamond}^E\leq \displaystyle \sup_{\Tr H_A\rho\leq E}\left[\Tr P^\bot_m\rho+\sqrt{\Tr P^\bot_m\rho}\right]\\\leq \displaystyle \bar{E}/\bar{E}_m+\sqrt{\bar{E}/\bar{E}_m},\;\; \textrm{where}\;\; P^\bot_m=I_A-P_m.
\end{array}
\end{equation}
So, by using Proposition 6 in \cite{SCT} and the change of variables $\,t\mapsto t/\varepsilon\,$ we obtain
$$
\begin{array}{rl}
|C_{\chi}(\Phi,H_A,E)-C_{\chi}(\Psi_m,H_A,E)|\, \leq f^{\alpha,E_c}_{C_{\chi}}(E,m|\,t)
\end{array}
$$
for any $\,t\in(0,\frac{1}{2}]$. This implies the assertion of the theorem for $C_*=C_{\chi}$, since the definition of the Holevo capacity and the implication (\ref{p-imp})
show that
$$
C_{\chi}(\Psi_m, H_{\!A},E)\leq C_{\chi}^m(\Phi, H_{\!A},E)\leq C_{\chi}(\Phi, H_{\!A},E).
$$
The assertions of the theorem for $C_*=C$ and $C_*=C_{\mathrm{ea}}$ are proved similarly by using
Proposition 6 and 7B in \cite{SCT}.

The assertions of the theorem for $C_*=Q$ is  proved by repeating the corresponding arguments from the proof of Theorem \ref{UFA}
with the use of Lemma \ref{b-lemma-el} below instead of Lemma \ref{b-lemma}. $\square$
\medskip

The following lemma is a version of Lemma \ref{b-lemma} in Section 3 adapted for energy limited channels.\smallskip

\begin{lemma}\label{b-lemma-el} \emph{Let  $\,\Pi_m:A\rightarrow A$ be the channel defined in Lemma \ref{b-lemma} and $\,\rho\shs$  a state  in $\,\S(\H^{\otimes n}_{A}\otimes\H_{R})$ such that $\,\sum_{k=1}^n\Tr H_A\rho_{A_k}\leq nE<+\infty$. If the Hamiltonian
$H_B$ of system $B$ satisfies condition (\ref{H-cond}) then
\begin{equation}\label{b-lemma-el+}
\left|I(B^n\!:\!R)_{\Phi^{\otimes n}\otimes\id_{R}(\rho)}-I(B^n\!:\!R)_{\Psi_m^{\otimes n}\otimes\id_{R}(\rho)}\right|\leq n f^{\alpha,E_c}_{Q}(E,m|\,p,t)
\end{equation}
for any $p>1$, $t\in(0,\frac{1}{2}]$ and any channel $\,\Phi\in\F_{\alpha,E_c}(A,B)$, where $\,\Psi_m=\Phi\circ\Pi_m$,  $f^{\alpha,E_c}_{Q}(E,m|\,p,t)$ is the quantity defined in Theorem \ref{UFA-el} and $\,\widehat{F}_{H_B}$ is any upper bound for the function $F_{H_B}$ with properties  (\ref{F-prop-1B}) and (\ref{F-prop-2B}).}\smallskip

\emph{If $\,\Tr H_A\rho_{A_k}\leq E\,$ for all $\,k=\overline{1,n}\,$ then (\ref{b-lemma-el+}) holds with $f^{\alpha,E_c}_{Q}(E,m|\,p,t)$ replaced by the quantity $f^{\alpha,E_c}_{C}(E,m|\,t)$ defined in Theorem \ref{UFA-el} for all $t\in(0,\frac{1}{2}]$.}
\end{lemma}

\smallskip

\emph{Proof.}  All the assertions of the lemma are easily derived from Lemma \ref{QCMI-CB} in the Appendix with trivial $C$ by using (\ref{ecd-up}). $\square$
\medskip

\begin{example}\label{exam+}
Let $A=B$ be the one-mode quantum oscillator with the frequency $\omega$. In this case $\{E^A_k=E^B_k=(k+1/2)\hbar\omega\}_{k\geq 0}$ and $F_{H_A}(E)=F_{H_B}(E)=g(E/\hbar\omega-1/2)$ \cite[Ch.12]{H-SCI}. The function defined in (\ref{q-osc}) with $\ell=1$, i.e. $\widehat{F}_{1,\omega}(E)\doteq\log(E/\hbar\omega+1/2)+1$ can be used in the role of the upper bound $\widehat{F}_{H_B}$.

Consider first the case $\alpha=1$, $E_c=0$. The set $\F_{1,0}(A,B)$ consists of channels not increasing the energy of a state, which can be called energy attenuators. The results of numerical calculations of $\,m_{C_*}(\varepsilon|\,1,0)$ for different values of the input energy bound $E$ are presented in the following tables corresponding to different values of the relative error $\varepsilon/F_{H_A}(E)$.\medskip

Table 3. The approximate values of $\,m_{C_*}(\varepsilon|\,\alpha,E_c)$ for $\varepsilon=0.1F_{H_A}(E)$,\\ \centerline{$\alpha=1$, $\;E_c=0$.}

\begin{tabular}{|c|c|c|c|c|}
  \hline
  $E/\hbar\omega$ & $m_{C_{\chi}}(\varepsilon|\,\alpha,E_c)$ & $m_C(\varepsilon|\,\alpha,E_c)$ & $m_{C_{\mathrm{ea}}}(\varepsilon|\,\alpha,E_c)$ & $m_Q(\varepsilon|\,\alpha,E_c)$ \\ \hline
  3 & $3.1\cdot10^{4}$ & $7.8\cdot10^{4}$ & $7.8\cdot10^{4}$ & $1.9\cdot10^{5}$ \\ \hline
  10 & $4.8\cdot10^{4}$ & $1.3\cdot10^{5}$ & $1.3\cdot10^{5}$ & $2.9\cdot10^{5}$ \\ \hline
  100 & $1.9\cdot10^{5}$ & $5.3\cdot10^{5}$ & $5.3\cdot10^{5}$ & $1.1\cdot10^{6}$ \\
  \hline
\end{tabular}
\medskip\pagebreak

Table 4. The approximate values of $\,m_{C_*}(\varepsilon|\,\alpha,E_c)$ for $\varepsilon=0.01F_{H_A}(E)$,\\ \centerline{$\alpha=1$, $\;E_c=0$.}

\begin{tabular}{|c|c|c|c|c|}
  \hline
  $E/\hbar\omega$ & $m_{C_{\chi}}(\varepsilon|\,\alpha,E_c)$ & $m_C(\varepsilon|\,\alpha,E_c)$ & $m_{C_{\mathrm{ea}}}(\varepsilon|\,\alpha,E_c)$ & $m_Q(\varepsilon|\,\alpha,E_c)$ \\ \hline
  3 & $5.6\cdot10^{6}$ & $1.3\cdot10^{7}$ & $1.3\cdot10^{7}$ & $3.1\cdot10^{7}$ \\ \hline
  10 & $8.5\cdot10^{6}$ & $2.0\cdot10^{7}$ & $2.0\cdot10^{7}$ & $4.7\cdot10^{7}$ \\ \hline
  100 & $3.3\cdot10^{7}$ & $8.3\cdot10^{7}$ & $8.3\cdot10^{7}$ & $1.8\cdot10^{8}$ \\
  \hline
\end{tabular}
\medskip

Comparing these results with the approximate values of $\,m_{C_*}(\varepsilon)$ presented in Tables 1 and 2 we see that the estimates of the $\varepsilon$-sufficient input dimensions given by Theorem \ref{UFA-el} for the class $\F_{1,0}(A,B)$ of energy attenuators  are significantly less than
the estimates of the $\varepsilon$-sufficient input dimensions given by Theorem \ref{UFA} for the class of all channel from the one-mode quantum oscillator to any other systems.

It is clear that $\,m_{C_*}(\varepsilon|\,\alpha,E_c)$ increases to $+\infty$ as either $\alpha$ or $E_c$ tends to $+\infty$. But  numerical calculations show that (in the case
when $A=B$ is the one-mode quantum oscillator) the rate of increasing of $\,m_{C_*}(\varepsilon|\,\alpha,E_c)$ is quite low for all the capacities. This is illustrated by the following tables corresponding to different values of the relative error $\varepsilon/F_{H_A}(E)$.
\medskip

Table 5. The approximate values of $\,m_{C_*}(\varepsilon|\,\alpha,E_c)$ for $\varepsilon=0.1F_{H_A}(E)$,\\ \centerline{$\alpha=10^6$, $\;E_c=10^6\hbar\omega$.}

\begin{tabular}{|c|c|c|c|c|}
  \hline
  $E/\hbar\omega$ & $m_{C_{\chi}}(\varepsilon|\,\alpha,E_c)$ & $m_C(\varepsilon|\,\alpha,E_c)$ & $m_{C_{\mathrm{ea}}}(\varepsilon|\,\alpha,E_c)$ & $m_Q(\varepsilon|\,\alpha,E_c)$ \\ \hline
  3 & $9.0\cdot10^{4}$ & $2.7\cdot10^{5}$ & $2.7\cdot10^{5}$ & $4.7\cdot10^{5}$ \\ \hline
  10 & $1.4\cdot10^{5}$ & $4.2\cdot10^{5}$ & $4.2\cdot10^{5}$ & $7.1\cdot10^{5}$ \\ \hline
  100 & $5.2\cdot10^{5}$ & $1.7\cdot10^{6}$ & $1.7\cdot10^{6}$ & $2.7\cdot10^{6}$ \\
  \hline
\end{tabular}
\medskip

Table 6. The approximate values of $\,m_{C_*}(\varepsilon|\,\alpha,E_c)$ for $\varepsilon=0.01F_{H_A}(E)$,\\ \centerline{$\alpha=10^6$, $\;E_c=10^6\hbar\omega$.}

\begin{tabular}{|c|c|c|c|c|}
  \hline
  $E/\hbar\omega$ & $m_{C_{\chi}}(\varepsilon|\,\alpha,E_c)$ & $m_C(\varepsilon|\,\alpha,E_c)$ & $m_{C_{\mathrm{ea}}}(\varepsilon|\,\alpha,E_c)$ & $m_Q(\varepsilon|\,\alpha,E_c)$ \\ \hline
  3 & $1.3\cdot10^{7}$ & $3.6\cdot10^{7}$ & $3.6\cdot10^{7}$ & $6.4\cdot10^{7}$ \\ \hline
  10 & $1.9\cdot10^{7}$ & $5.4\cdot10^{7}$ & $5.4\cdot10^{7}$ & $9.7\cdot10^{7}$ \\ \hline
  100 & $7.2\cdot10^{7}$ & $2.1\cdot10^{8}$ & $2.1\cdot10^{8}$ & $3.6\cdot10^{8}$ \\
  \hline
\end{tabular}
\medskip

Comparing Tables 5 and 6 with the Tables 3 and 4 shows that the change of the parameters $\,\alpha: 1\rightarrow 10^6\,$ and $\,E_c: 0\rightarrow 10^6 \hbar\omega\,$
does not lead to significant growth of the $\varepsilon$-sufficient input dimensions for all the capacities.

\medskip

\end{example}

\section{Uniform continuity of basic capacities of energy-constrained channels with respect to the strong convergence topology}

Real physical channels are always prepared with a finite accuracy. So, in study of their capacities we should be able to estimate variations of the capacities caused by all possible perturbations of a channel. In other words, we have to quantitatively analyse continuity of quantum channel capacities as functions of a channel  with respect to appropriate topology (convergence) on the set of all channels.

In finite dimensions this problem is solved by Leung and Smith who obtained in \cite{L&S} (uniform) continuity bounds for basic capacities of quantum channels with finite-dimensional output with respect to the distance between quantum channels generated by the diamond norm (\ref{d-norm}).

Speaking about generalizations of the Leung-Smith results to energy-constrained infinite-dimensional channels we have to choose appropriate metric on the set of quantum channels, since the diamond-norm distance can not properly describe all physical perturbations of infinite-dimensional channels (this is illustrated by the examples of channels with close physical parameters  having the diamond-norm distance equal to $2$ \cite{W-EBN}).

Mathematically, the drawback of the  diamond-norm distance in infinite-dimensions
follows from  Theorem 1 in \cite{Kr&W} stating that  the closeness of two quantum channels in the diamond-norm distance means \emph{the operator norm} closeness of the corresponding Stinespring isometries. To take into account deformations of the Stinespring
isometry in the strong operator topology one can consider the \emph{strong convergence topology} on the set of quantum channels defined by the family of seminorms $\Phi\mapsto\|\Phi(\rho)\|_1, \rho\in\S(\H_A)$ \cite{SCT}. The strong convergence of a sequence of channels $\Phi_n$ to a channel $\Phi_0$  means that
$$
\lim_{n\rightarrow\infty}\Phi_n(\rho)=\Phi_0(\rho)\,\textup{ for all }\rho\in\S(\H_A).
$$
The separability of the set $\S(\H_A)$ implies
that the strong convergence topology on the set of quantum channels is metrisable
(can be defined by some metric). Moreover, it is shown in \cite{SCT} that this topology is generated by any of the energy-constrained diamond norms
(\ref{ecd}) provided the operator $H_A$ has discrete spectrum $\{E^A_k\}_{k\geq0}$ of finite multiplicity such that $\,E^A_k\rightarrow+\infty$ as $k\rightarrow+\infty$.

In \cite{SCT,W-EBN} continuity bounds for basic capacities of infinite-dimensional energy-constrained channels with respect to the energy-constrained diamond norms (\ref{ecd}) are obtained under the condition of boundedness of the energy amplification factor of these channels. The continuity bound for the entanglement-assisted capacity $C_{\mathrm{ea}}$ obtained in \cite{SCT} holds for arbitrary quantum channels and hence implies
uniform continuity of this capacity on the set of \emph{all quantum channels} with respect to the strong convergence topology provided the input Hamiltonian $H_A$ satisfies condition (\ref{H-cond}). The finite-dimensional approximation theorem makes it possible to obtain similar result for other basic capacities under slightly stronger condition on $H_A$.\smallskip

\begin{theorem}\label{cap-uc} \emph{If the Hamiltonian $H_A$ of input system $A$ satisfies condition (\ref{H-cond+}) then  for any $\,E>E^A_0$ all the functions
$$
\Phi\mapsto C_*(\Phi,H_A,E),\quad C_*=C_{\chi}, C, Q, C_{\mathrm{p}},
$$
are uniformly continuous on the set of all channels from $A$ to arbitrary system $B$ with respect to the strong convergence topology. Quantitatively, if $\,\Phi$ and $\,\Psi$ are any channels from $A$ to $B$ such that $\,\frac{1}{2}\|\Phi-\Psi\|_{\diamond}^E\leq\varepsilon$ then
\begin{equation}\label{HC-CB-be}
|\shs C_*(\Phi,H_A,E)-C_*(\Psi,H_A,E)|\leq v_{C_*}(\varepsilon,E),\quad C_*=C_{\chi}, C, Q, C_{\mathrm{p}},
\end{equation}
where $v_{C_*}(\varepsilon,E)$ is a function vanishing as $\,\varepsilon \rightarrow0^+$ for any $E>E^A_0$ defined for each of the capacities by the formulas
$$
v_{C_{\chi}}(\varepsilon,E)=\min_{m\in\mathbb{N}_*}\!\left[\sqrt{k(m)\varepsilon}\log (2m)+2g\!\left(\!\sqrt{k(m)\varepsilon}\right)+2f_{C_{\chi}}(E,m)\right]\!,\;\,
$$
$$
v_{C}(\varepsilon,E)=\min_{m\in\mathbb{N}_*}\!\left[2\sqrt{k(m)\varepsilon}\log (2m)+2g\!\left(\!\sqrt{k(m)\varepsilon}\right)+2f_{C}(E,m)\right]\!,
$$
$$
v_{Q}(\varepsilon,E)=\min_{m\in\mathbb{N}_*}\!\left[2\sqrt{k(m)\varepsilon}\log (2m)+2g\!\left(\!\sqrt{k(m)\varepsilon}\right)+2f_{Q}(E,m)\right]\!,
$$
$$
\;\,v_{C_{\mathrm{p}}}(\varepsilon,E)=\min_{m\in\mathbb{N}_*}\!\left[4\sqrt{k(m)\varepsilon}\log (2m)+4g\!\left(\!\sqrt{k(m)\varepsilon}\right)+2f_{C_{\mathrm{p}}}(E,m)\right]\!,\;\,
$$
where $\,\mathbb{N}_*\doteq\{m\in\mathbb{N}\,|\, E^A_m\geq 16E-15E^A_0\}$, the functions $f_{C_{\chi}}$, $f_{C}$, $f_{Q}$ and $f_{C_{\mathrm{p}}}$ are defined in Theorem \ref{UFA} and $\,k(m)=2(E^A_m-E^A_0)/(E-E^A_0)$.}\footnote{$E^A_m$ is the $m$-th eigenvalue of $H_A$ (taking the multiplicity into account).}
\end{theorem}
\medskip

\emph{Proof.} The first assertion of the theorem follows from continuity bounds (\ref{HC-CB-be}), since the energy-constrained diamond norm $\|\cdot\|^{E}_{\diamond}$ with any $E>E^A_0$ generates the strong convergence topology on the set of all quantum channels from $A$ to $B$ by Proposition 3 in \cite{SCT}.\smallskip

For given natural $m$ let $\Phi_m$ and $\Psi_m$ be the restrictions of the channels $\Phi$ and $\Psi$ to the set $\S(\H^m_A)$. By repeating the arguments from the proof of Proposition 5 in \cite{CID} one can show that
\begin{equation}\label{m-cb}
\begin{array}{l}
|\shs C_{\chi}(\Phi_m,H^m_A,E)-C_{\chi}(\Psi_m, H^m_A, E)|\leq \epsilon
\log (2m) +2g(\epsilon),\\\\
\left|\shs C(\Phi_m,H^m_A,E)-C(\Psi_m,H^m_A,E)\right|\leq 2\epsilon
\log (2m)+2g(\epsilon),\\\\
\left|\shs Q(\Phi_m,H^m_A,E)-Q(\Psi_m,H^m_A,E)\right|\leq 2\epsilon
\log (2m)+2g(\epsilon),\\\\
\left|\shs C_{\mathrm{p}}(\Phi_m,H^m_A,E)-C_{\mathrm{p}}(\Psi_m,H^m_A,E)\right|\leq 4\epsilon
\log (2m) +4g(\epsilon),
\end{array}
\end{equation}
where $\,\epsilon=\,\|\Phi_m-\Psi_m\|^{1/2}_{\diamond}$ and $H^m_A=H_AP_m$ (here $P_m$ is the projector onto $\H_A^m$). Since $\|\Phi_m-\Psi_m\|_{\diamond}=\sup\limits_{\omega_A\in\S(\H^m_A)}\|(\Phi-\Psi)\otimes\id_R(\omega)\|_1$,
by noting that $\Tr H_A\rho\leq E^A_m$ for any $\rho\in \S(\H^m_A)$ and by using monotonicity and concavity of the function $E\mapsto \|\Phi\|^E_{\diamond}$ (proved in \cite{W-EBN}) we obtain
$$
\epsilon^2\leq\|\Phi-\Psi\|^{E^A_m}_{\diamond}\leq \textstyle\frac{1}{2}k(m)\|\Phi-\Psi\|^{E}_{\diamond}\leq k(m)\varepsilon.
$$

If $C_*$ is one of the capacities  $C_{\chi}, C, Q$ and $C_{\mathrm{p}}$ then
$$
\begin{array}{ccc}
|\shs C_*(\Phi,H_A,E)-C_*(\Psi,H_A,E)|\leq|\shs C_*(\Phi_m,H_A,E)-C_*(\Psi_m,H^m_A,E)|\\\\+|\shs C_*(\Phi,H_A,E)-C_*(\Phi_m,H^m_A,E)|+|\shs C_*(\Psi,H^m_A,E)-C_*(\Psi_m,H^m_A,E)|.
\end{array}
$$
Thus, the continuity bounds in the theorem follow from continuity bounds (\ref{m-cb}) and Theorem \ref{UFA}.

The vanishing of all the functions $v_{C_*}(\varepsilon,E)$ as $\varepsilon\rightarrow0^+$ follows from the
vanishing of  the functions $f_{C_*}(E,m)$ as $m\rightarrow+\infty$. $\square$

\medskip

\begin{remark}\label{th-2-r}
Continuity bounds (\ref{HC-CB-be}) are universal (valid for any channels) but they give too rough estimates for variations of the capacities because of the low decreasing rate of the functions $f_{C_*}(E,m)$ as $m\rightarrow+\infty$.
So, dealing with quantum channels produced in a physical experiment it is reasonable to use the continuity bounds for basic capacities depending on the energy-constrained diamond norm distance obtained in \cite{SCT,W-EBN} for classes of channels with bounded energy amplification factor.
\end{remark}

\section*{Appendix}

The following lemma is the QCMI-version of Lemma 7 in \cite{W-EBN}.\footnote{The proof of this lemma differs from the proof of Lemma 7 in \cite{W-EBN} (containing similar continuity bound for the conditional entropy) by the way of splitting of $\{1,2,...,n\}$ into the sets $N_1$ and $N_2$. This makes the resulting continuity bound more accurate in the case of logarithmic growth of $F_{H_B}$ (in particular, when $B$ is a multi-mode quantum oscillator).}  \smallskip

\begin{lemma}\label{QCMI-CB} \emph{Let $\,\Phi$ and $\,\Psi$ be channels from $A$ to $B$ satisfying condition (\ref{elc}), $C,D$ any systems, $n\in\mathbb{N}$ and $\,\rho\shs$  a state  in $\,\S(\H^{\otimes n}_{A}\otimes\H_{CD})$ such that $\,\sum_{k=1}^n\Tr H_A\rho_{A_k}\leq nE<+\infty$. If $\,\frac{1}{2}\|\Phi-\Psi\|_{\diamond}^E\leq\varepsilon$ then\footnote{$\|\cdot\|_{\diamond}^E$ is the energy-constrained diamond norm defined in (\ref{ecd}).}
\begin{equation}\label{QCMI-CB+}
\begin{array}{l}
(1/n)\left|I(B^n\!:\!D|C)_{\Phi^{\otimes n}\otimes\id_{CD}(\rho)}-I(B^n\!:\!D|C)_{\Psi^{\otimes n}\otimes\id_{CD}(\rho)}\right|\\\\
\leq\displaystyle(4t+2r(t,\varepsilon))\widehat{F}_{H_{B}}\!\left(\!\frac{E_p}{t}\right)+2g\!\left(r(t,\varepsilon)\right)+4h_2(t)+
\frac{2}{p}\widehat{F}_{H_{B}}(E_p)
\end{array}
\end{equation}
for any $\,p>1$ and $\,t\in(0,\frac{1}{2}]$, where $E_p=\alpha p E+E_c$, $r(t,\varepsilon)=\frac{\varepsilon+t/2}{1-t}$ and $\,\widehat{F}_{H_B}$ is any upper bound for the function $F_{H_B}$ with properties  (\ref{F-prop-1B}) and (\ref{F-prop-2B}).}\smallskip

\emph{If $\,\Tr H_A\rho_{A_k}\leq E\,$ for all $\,k=\overline{1,n}\,$ then (\ref{QCMI-CB+}) holds with $p=1$ without the last term in the right hand side.} \smallskip
\end{lemma}\smallskip

\emph{Proof.} Denote by $\Delta^n(\Phi,\Psi,\rho)$ the left hand side of (\ref{QCMI-CB+}). By the proof of Proposition 3B in \cite{CHI} (based on the Leung-Smith telescopic method), we have
$$
n\Delta^n(\Phi,\Psi,\rho)\leq \sum_{k=1}^{n}|I(B_k\!:\!D|X)_{\sigma_k}-I(B_k\!:\!D|X)_{\sigma_{k-1}}|,
$$
where $X=B_1...B_{k-1}B_{k+1}...B_nC$ and $\sigma_k=\Phi^{\otimes k}\otimes\Psi^{\otimes (n-k)}\otimes\id_{CD}(\rho)$, $k=0,1,...,n$. The proof of Proposition 3B in \cite{CHI} also implies
\begin{equation}\label{t-ineq}
\|\sigma_k-\sigma_{k-1}\|_1\leq \sup\left\{\|(\Phi-\Psi)\otimes\id_R(\omega)\|_1\,|\;\omega_{A}=\rho_{A_k}\right\}\leq \|\Phi-\Psi\|_{\diamond}^{x_k},
\end{equation}
where $x_k=\Tr H_A\rho_{A_k}$.\smallskip

Since $[\sigma_k]_{B_k}=\Phi(\rho_{A_k})$ and $[\sigma_{k-1}]_{B_k}=\Psi(\rho_{A_k})$, we have
\begin{equation}\label{energy-b}
\Tr H_B[\sigma_k]_{B_k}, \Tr H_B[\sigma_{k-1}]_{B_k}\leq\alpha x_k + E_c.
\end{equation}

Let $N_1$ be the set of indexes $k$ for which $x_k\leq p E$ and $N_2=\{1,..,n\}\setminus N_1$. Thus,
$$
n\Delta^n(\Phi,\Psi,\rho)\leq \sum_{k\in N_1}\!D_k+\sum_{k\in N_2}\!D_k,\quad D_k=|I(B_k\!:\!D|X)_{\sigma_k}-I(B_k\!:\!D|X)_{\sigma_{k-1}}|.
$$

For each $k\in N_1$  Proposition 2 in \cite{CHI} along with (\ref{t-ineq}) and (\ref{energy-b}) imply
$$
\!D_k\leq \displaystyle\left(4\varepsilon_kt_k+\frac{2\varepsilon_k+\varepsilon_kt_k}{1-\varepsilon_kt_k}\right)\!\widehat{F}_{H_B}\!\!\left(\frac{\alpha x_k+E_c}{\varepsilon_k t_k}\right)
+2g\!\left(\frac{\varepsilon_k+\varepsilon_kt_k/2}{1-\varepsilon_kt_k}\right)+4h_2(\varepsilon_k t_k),
$$
for any $\,t_k\in(0,\frac{1}{2\varepsilon_k}]$, where $\varepsilon_k=\frac{1}{2}\|\Phi-\Psi\|^{x_k}_{\diamond}$. By choosing free parameters $t_k$ such that $\,\varepsilon_kt_k=t\,$ for all $k\in N_1$ we obtain
$$
\!\!\!\begin{array}{rl}
\displaystyle
\sum_{k\in N_1}D_k \leq & \!\!\!\displaystyle \sum_{k\in N_1}\!\left(\!\left(4t+\frac{2\varepsilon_k+t}{1-t}\right)\!\widehat{F}_{H_B}\!\!\left(\frac{\alpha x_k+E_c}{t}\right)
+2g\!\left(\frac{\varepsilon_k+t/2}{1-t}\right)\!\right)+4n_1h_2(t)\\\\ \leq & \!\!\displaystyle
n_1\!\left(4t+\frac{2\bar{\varepsilon}_1+t}{1-t}\right)\!\widehat{F}_{H_B}\!\!\left(\frac{\alpha pE +E_c}{t}\right)
+2n_1g\!\left(\frac{\bar{\varepsilon}_1+t/2}{1-t}\right)+4n_1h_2(t),
\end{array}
$$
where $n_1=\sharp (N_1)$ and $\bar{\varepsilon}_1\doteq n_1^{-1}\sum_{k\in N_1}\varepsilon_k$. The last inequality follows from monotonicity of the function $\widehat{F}_{H_B}$ (since $x_k\leq pE$ for all $k\in N_1$) and  concavity  of the function $g(x)$.

By using  monotonicity and concavity of the function $E\mapsto \|\Phi\|^E_{\diamond}$ (proved in \cite{W-EBN}) it is easy to show that $\bar{\varepsilon}_1\leq \frac{1}{2}\|\Phi-\Psi\|_{\diamond}^E\leq\varepsilon$.  So, by monotonicity of $g(x)$ we have
$$
n^{-1}\sum_{k\in N_1}D_k\leq \left(4t+\frac{2\varepsilon+t}{1-t}\right)\!\widehat{F}_{H_B}\!\!\left(\frac{\alpha pE +E_c}{t}\right)
+2g\!\left(\frac{\varepsilon+t/2}{1-t}\right)+4h_2(t).
$$

For each $k\in N_2$ upper bound (\ref{CMI-UB}), nonnegativity of QCMI and inequalities (\ref{energy-b}) imply
$$
D_k\leq 2\max\{H([\sigma_k]_{B_k}), H([\sigma_{k-1}]_{B_k})\}\leq 2\widehat{F}_{H_B}(\alpha x_k + E_c).
$$
So, by concavity of $\widehat{F}_{H_B}$ we have
$$
\sum_{k\in N_2}D_k\leq 2\sum_{k\in N_2}\widehat{F}_{H_B}(\alpha x_k + E_c)\leq 2n_2\widehat{F}_{H_B}(\alpha X_2 + E_c),
$$
where $n_2=\sharp (N_2)$ and $X_2=n_2^{-1}\sum_{k\in N_2}x_2$. Since $\sum_{k\in N_2}x_k\leq nE$ and $x_k> pE$ for all $k\in N_2$, we have $X_2\leq nE/n_2$ and $n_2/n\leq 1/p$. By using monotonicity of $\widehat{F}_{H_B}$ and applying Lemma \ref{GWL} to the concave nonnegative function $x\mapsto\widehat{F}_{H_B}(\alpha x + E_c)$ on $\mathbb{R}_+$ we obtain
$$
n^{-1}\sum_{k\in N_2}D_k\leq 2(n_2/n)\widehat{F}_{H_B}(\alpha (n/n_2) E + E_c)\leq (2/p)\widehat{F}_{H_B}(\alpha p E + E_c).
$$
This and the above estimate for $n^{-1}\sum_{k\in N_1}D_k$ imply (\ref{QCMI-CB+}).\smallskip

The last assertion of the lemma follows from the above arguments with $p=1$, since in this case the set $N_2$ is empty. $\square$ \medskip

{\bf Acknowledgments.}  I am grateful to the  participants of the workshop "Recent advances in continuous variable quantum information theory", Barcelona, April, 2016  for the stimulating  discussion. I am grateful to A.Winter for sending me a preliminary version of the paper \cite{W-EBN}  used in this work. I am also grateful to A.S.Holevo and G.G.Amosov for useful comments and
to M.M.Wilde for valuable communication concerning  capacities of infinite-dimensional channels with energy constraints. Special thanks to Yu.V.Andreev and L.V.Kuzmin for the help with MathLab.\medskip

The research is funded by the grant of Russian Science Foundation
(project No 14-21-00162).

\end{document}